\documentclass{elsart5p}
\usepackage{graphicx,natbib,amssymb}
\journal{New Astronomy}
\newcommand{\ubv}{\,{\it UBV~}}
\newcommand{\kpc}{\,{\rm kpc}}

\newcommand{\cmt}{\,cm$^{-3}$}
\newcommand{\cmd}{\,cm$^{-2}$}
\newcommand{\myr}{\,$M_{\odot}\,{\rm yr}^{-1}$}
\newcommand{\es}{$\,\rm erg\,s^{-1}$}
\newcommand{\ecs}{$\,\rm erg\,cm^{-2}\,s^{-1}$}

\newcommand{\ha}{H$\alpha$}

\newcommand{\ro}{\,$R_{\odot}$}
\newcommand{\mo}{\,$M_{\odot}$}
\newcommand{\lo}{\,$L_{\odot}$}
\def\I{\rm {\scriptsize I}}

\begin{document}
\begin{frontmatter}
\title{Multiwavelength modelling the SED of supersoft X-ray sources \\[2mm]
I. The method and examples}
%

\author{A.~Skopal\thanksref{fn1}, \thanksref{fn2}}

\thanks[fn1]{E-mail: skopal@ta3.sk}
\thanks[fn2]{Visiting Astronomer: Astronomical Institut, Bamberg}
\thanks[fn3]{doi:?}

\address{Astronomical Institute, Slovak Academy of Sciences,
         059\,60 Tatransk\'{a} Lomnica, Slovakia \\[1mm]
        {\rm Received 15 July 2013; accepted 17 October 2013}
        }

\vspace*{5mm}

{\small 
\hspace*{-15cm}
{\rm H I G H L I G H T S}\\
\begin{itemize}
\item
New method of modelling the spectra of supersoft X-ray sources is introduced. 
\item
The method analyses the X-ray/near-IR spectral energy distribution. 
\item
The multiwavelength approach overcomes problems of modelling only the X-ray data. 
\item
More trustworthy fundamental parameters are derived. 
\end{itemize}
}

\vspace*{-7mm}

\begin{abstract}
Radiation of supersoft X-ray sources (SSS) dominates both the 
supersof X-ray and the far-UV domain. A fraction of their 
radiation can be reprocessed into the thermal nebular emission, 
seen in the spectrum from the near-UV to longer wavelengths. 
In the case of symbiotic X-ray binaries (SyXBs) a strong 
contribution from their cool giants is indicated 
in the optical/near-IR. 
In this paper I introduce a method of multiwavelength modelling 
the spectral energy distribution (SED) of SSSs from the supersoft 
X-rays to the near-IR with the aim to determine the physical 
parameters of their composite spectra. 
The method is demonstrated on two extragalactic SSSs, the SyXB 
RX\,J0059.1-7505 (LIN~358) in the Small Magellanic Cloud (SMC), 
RX\,J0439.8-6809 in the Large Magellanic Cloud (LMC) and two 
Galactic SSSs, the classical nova RX\,J2030.5+5237 (V1974~Cyg) 
during its supersoft phase and the classical symbiotic star 
RX\,J1601.6+6648 (AG~Dra) during its quiescent phase. 
The multiwavelength approach overcomes the problem of the mutual 
dependence between the temperature, luminosity and amount of 
absorption, which appears when only the X-ray data are fitted. 
Thus, the method provides an unambiguous solution. 
It was found that selection of the model (a blackbody or an 
atmospheric model) is not of crucial importance in fitting 
the global X-ray/IR SED. 
The multiwavelength modelling of the SED of SSSs is essential 
in determining their physical parameters. 
\end{abstract} 
\end{frontmatter}
\section{Introduction}

SSSs were first detected in the Magellanic Clouds with 
the \textsl{EINSTEIN} and \textsl{EXOSAT} satellites 
\citep[e.g.][]{long+81,sm81,pakull+85}. Later observations 
with \textsl{ROSAT} verified their supersoft nature and showed 
that SSSs do not emit significantly at 
energies above $\sim0.5$\,keV \citep[e.g.][]{greiner+91}. 
Typical blackbody parameters of SSSs are 
a temperature of $3-5\times 10^{5}$\,K and an effective radius 
of $1-3\times 10^{9}$\,cm, which suggests their luminosities 
to be as high as $\sim 10^{38}$\es. First thoughts about 
the nature of the SSSs included accretion 
onto black holes \citep[][]{cowley+90} and neutron stars 
accreting above the Eddington rate \citep[][]{greiner+91}. 
\cite{vdh+92} discussed difficulties with the neutron star/black 
hole model, and proposed the possibility that the supersoft 
X-ray emission is the result of steady nuclear burning of 
hydrogen accreted onto the surface of a massive white dwarfs (WDs). 
At present, this scenario represents the favoured model explaining 
the high energy output produced by these objects. Various 
phenomena associated with mass-accreting WDs in SSSs were 
reviewed by \cite{kato10}. 

SSSs are understood as interacting binary systems 
consisting of an accreting compact object and a low-mass 
($\le 2 M_{\odot}$) main-sequence or slightly evolved 
late-type star. In special cases, the donor star can also 
be an M-type giant. The latter sources are often called as SyXBs. 
They were identified by finding a red giant counterpart to 
the primarily detected X-ray source 
\citep[e.g.][]{davidsen,masseti+06,nespoli+10}. 
On the other hand, many classical symbiotic stars have also 
been detected as X-ray sources 
\citep[e.g.][]{muerset+97,wheatley,stute+11}, as a natural 
consequence of their hot component properties, 
\citep[$T_{\rm h} \gtrsim 10^5$\,K, $L_{\rm h} 
\sim 10^3 - 10^4\,L_{\odot}$, e.g.][]{m+91,sk05}. 
Transient SSSs emerge also during nova outbursts, when 
the radius of the expanding WD pseudophotosphere gradually 
shrinks, radiates at higher temperature, and thus shifts 
the maximum of its energy distribution from the optical to 
soft X-rays \citep[][]{gall+code74}. 

Determination of physical parameters of SSSs has been usually 
based solely on analyzing the X-ray data, which, however, cover 
only a very small fraction of their total spectrum. As a result, 
a good fit to the X-ray data was often achieved with very 
different sets of parameters and/or models. 
For example, \cite{heise94} obtained a satisfactory fit 
($\chi^2_{\rm red} < 1$) to the \textsl{ROSAT} data of a stable 
SSS, SMP SMC 22, with a blackbody and an atmospheric model. 
However, the luminosity of their blackbody fit was more 
than a factor of 10 higher than that given by the atmospheric 
model. Recently, \cite{mereghetti+10} studied this object with 
\textsl{XMM-Newton}, and also fitted successfully the X-ray 
data with both the blackbody and a NLTE atmospheric model. 
They did not find a significant difference between luminosities 
from both models, but the temperature of their preferred 
atmospheric model was only $1.54 \times 10^{5}$\,K, in contrast 
to $4.4 \times 10^{5}$\,K of the Heise et al. model. 
Satisfactory fits to X-ray observations, corresponding to very 
wide range of possible parameters, were found by many 
authors \citep[e.g.][]{greiner+97,asai+98,ness+08,sturm+11}. 
Such the ambiguity in the inferred parameters is probably caused 
by fitting only a very small part of the total SSS spectrum. 

Accordingly, in this paper I propose a method to determine the 
fundamental, $L,R,T$, and $N_{\rm H}$ parameters of SSSs 
by {\em multiwavelength} modelling their global SED throughout 
a very large, X-ray to near-IR, spectral range (Sect.~2). 
In Sect.~3 I apply the method to selected objects and introduce 
the results. Their discussion and summary are found in 
Sects.~4 and 5, respectively.

\section{Multiwavelength model SED}

\subsection{Composite spectrum}

Depending on the nature of a SSS, there can be a certain and 
variable amount of the circumstellar material (CSM) in the 
system, which is subject to ionization by the SSS radiation. 
For example, during the nebular stage of nova outburst, the 
ejected material reprocesses a fraction of the hot WD's radiation 
into the nebular emission. In the case of SyXBs, the nebular 
component of radiation is produced also by the ionized part 
of the neutral wind from the cool giant. 
Thus the nebular component has to be included in the 
multiwavelength modelling the SED. In addition, the cool giant 
in SyXBs dominates the near-IR with a significant contribution 
to the optical. As a result, the composite spectrum emitted by 
a SSS, $F(\lambda)$, can be generally expressed as 
a superposition of these three basic radiative components, 
%
%
\begin{equation}
  F(\lambda) = F_{\rm h}(\lambda) + F_{\rm n}(\lambda) + 
               F_{\rm g}(\lambda). 
\label{eq:fsss}
\end{equation}
%
$F_{\rm h}(\lambda)$ is the flux produced by the hot 
{\em stellar} (pseudo)photo\-sphere of a SSS (i.e. its central 
star and/or disk), $F_{\rm n}(\lambda)$ is the {\em nebular} 
component from the ionized CSM (i.e. that produced by thermal 
plasma) and $F_{\rm g}(\lambda)$ is the contribution 
from the cool giant. 

To achieve the aim of this paper, it means to disentangle 
the observed composite spectrum into its individual components, 
i.e. to determine their physical parameters. 
However, the radiation observed at the Earth have to be corrected 
for the attenuation by the ISM and CSM, placed between the 
source and the observer. 

\subsection{Corrections of the observed light}

Assuming that no emission occurs on the path from the source to 
the observer, we can use the simplest solution of the radiative 
transfer equation to correct the observed SSS fluxes as 
%
%
\begin{equation}
 F_{\rm h}^{\rm obs}(\lambda) = \theta_{\rm h}^2 
           F_{\rm h}(\lambda)\,e^{-\tau_{\lambda}},
\label{eq:ftaux}
\end{equation}
where $F_{\rm h}(\lambda)$ is the spectrum emitted by the SSS, 
its angular radius $\theta_{\rm h}$ scales it to that observed at 
the Earth ($F_{\rm h}^{\rm obs}(\lambda)$) and the optical depth 
$\tau_{\lambda}$ attenuates it along the line of sight. 
In the X-ray domain we consider attenuation of the light by 
{\rm b--f} absorptions, while for wavelengths $> 912$\,\AA\ 
we consider selective extinction on dust particles as the only 
absorber within the interstellar matter (ISM). 
In the former case, $\tau_{\lambda}$ is given by the absorption
cross-section per atom, $\sigma_i(\lambda)$, of the element,
$i$, and the total number of atoms on the line of sight, 
$\sum_i n_i$. Thus, 
$\tau_{\lambda} = \sum_i n_i \sigma_i(\lambda) = 
                  \sum_i a_i\sigma_i(\lambda) N_{\rm H}$, 
where $a_i$ is the relative abundance and $N_{\rm H}$ 
the total hydrogen column density [\cmd] (i.e. through the 
ISM and CSM, see below). 
Then the optical depth, caused by the {\rm b--f} 
absorptions within the X-ray domain, can be written in 
a convenient form as 
%
%
\begin{equation}
  \tau_{\rm x}(\lambda) = \sigma_{\rm x}(\lambda) N_{\rm H}, 
\label{eq:taux}
\end{equation}
%
where $\sigma_{\rm x}(\lambda)$ [cm$^{2}$] is the total 
cross-section for photoelectric absorption per hydrogen atom 
\citep[e.g.][]{crudd+74}. To correct the observed X-ray fluxes 
for absorptions I used the {\em tbabs} absorption model for 
ISM composition with abundances given by \cite{wilms+00} 
(e.g. $\log(A_{\rm OI})+12 = 8.69$). 

The minimum of $N_{\rm H}$ is given by its ISM component, 
$N_{\rm H}$(ISM). The CSM component, $N_{\rm H}$(CSM), can 
be variable, depending on the activity of a SSS and/or its 
composition. For example, an increase of the mass-loss 
rate from a SSS leads to a decrease of the supersoft spectrum 
due to the {\rm b--f} transitions, while the near-UV/optical 
spectrum brighten up throughout the {\rm f--b} and {\rm f--f} 
transitions, and vice versa. This effect is measured in some 
SSSs (see Paper\,II). A nice example is AG~Dra, which shows 
a strict anticorrelation between the supersoft X-ray and 
near-UV/optical fluxes \citep[][]{greiner+97,gon+08,sk+09}. 
If a SSS contains a strong source of the neutral hydrogen (SyXBs), 
the amount of {\rm b--f} absorptions on the line of sight can vary 
with the orbital phase, when viewing the hot accreting source 
throughout a different amount of the neutral stellar wind from 
the giant. An example here is SMC\,3 \citep[][]{kahabka04} and 
the symbiotic binary 4~Dra \citep[see Fig~6 of][]{wheatley}. 
%
The above-mentioned examples imply that the amount of absorption 
between the observer and a SSS can be {\em variable} due to its 
CSM component. Thus in the modelling the SED one 
has to consider that 
%
%
\begin{equation}
    N_{\rm H} = N_{\rm H}({\rm ISM}) + N_{\rm H}({\rm CSM}). 
\label{eq:nh}
\end{equation}

Photons that are not capable of ionizing hydrogen are attenuated 
predominantly by scattering on the dust grains in the ISM. 
In this case the light attenuation is given by the extinction 
curve, $k_{\lambda}$ \citep[e.g.][]{c+89}, parametrized with 
the colour excess, $E_{\rm B-V}$, to the object. This 
correction to the observed fluxes, $F^{\rm obs}(\lambda)$, 
for $\lambda > 912$\,\AA\ can be expressed as 
%
%
\begin{equation}
 F^{\rm obs}(\lambda) = F^{\rm der}(\lambda) 
            10^{-0.4\,R_{\rm V}\,k_{\lambda}\,E_{\rm B-V}}, 
\label{eq:extin}
\end{equation}
where $F^{\rm der}(\lambda)$ is the so-called `dereddened' 
flux and the ratio of total to selective extinction 
$R_{\rm V} = 3.1$ \citep[e.g.][]{wegner}. 

In the following sections, I will briefly introduce the radiative 
components of Eq.~(\ref{eq:fsss}) as observed at the Earth, and 
will outline principles of the method of disentangling 
the composite continuum. 

\subsection{The hot stellar continuum}

It is understood that the radiation from SSSs is produced 
by their (pseudo)photospheres, and thus it is of 
stellar nature. A good example is the supersoft X-ray radiation 
generated by the WD photosphere as measured from novae during 
their SSS phases \citep[e.g.][]{gall+code74,krautter+96}. 
Also the supersoft X-ray radiation from AG~Dra (the strongest 
source among classical symbiotics) was associated with the hot 
star photosphere \citep[][]{greiner+97,viotti+05}, and later 
confirmed with the model SED by \cite{sk+09}. 

For the sake of simplicity and with respect to the current 
modelling the low-resolution supersoft X-ray data, the observed 
fluxes from the hot stellar source (i.e. the SSS) can be 
approximated with the blackbody radiation at a temperature 
$T_{\rm h}$, which is attenuated by {\rm b--f} absorptions in 
the X-ray domain. 
Then, according to Eqs.~(\ref{eq:ftaux}) and (\ref{eq:taux}), 
the observed fluxes from the SSS can be fitted with 
%
%
\begin{equation}
  F^{\rm obs}_{\rm h}(\lambda) =
       \theta_{\rm h}^2 \pi B_{\lambda}(T_{\rm h})\,
       e^{-\sigma_{\rm x}(\lambda)\,N_{\rm H}},
\label{eq:fhot}
\end{equation}
where the angular radius of the SSS, 
$\theta_{\rm h} = R_{\rm h}^{\rm eff}/d$, is given by its 
effective radius, $R_{\rm h}^{\rm eff}$ (i.e. the radius of 
a sphere with the same luminosity) and the distance $d$. 
Fitting parameters here are $\theta_{\rm h}$, $T_{\rm h}$
and $N_{\rm H}$, which define the luminosity of the SSS as 
$L_{\rm h} = 4\pi d^2 \theta_{\rm h}^2 \sigma T_{\rm h}^4$. 
If required, the blackbody fluxes in Eq.~(\ref{eq:fhot}) can 
be replaced by a more sophisticated model of a hot stellar 
atmosphere or accretion disk. 
%

\subsubsection{$N_{\rm H}$ from the Rayleigh scattering}

Rayleigh scattering describes the absorption process, when an 
atom is excited to the {\em intermediate} state and is 
immediately stabilized by a transition to the same true 
bound state, re-emitting a photon of the same wavelength 
\citep[see Fig.~1 of][]{nussb+89}. 
The effect is best observable for transitions from the ground 
state of hydrogen atom, i.e. around the lines of the hydrogen 
Lyman series, where it can be identified as a strong attenuation 
of the continuum. Its strength is therefore determined by the 
column density of neutral hydrogen between the emitting source 
and the observer, causing the optical depth in the continuum, 
%
%
\begin{equation}
  \tau_{\rm ray}(\lambda) = \sigma_{\rm ray}(\lambda) N_{\rm H}, 
\label{eq:ray}
\end{equation}
where $\sigma_{\rm ray}(\lambda)$ [cm$^{2}$] is the Rayleigh 
cross-section for scattering by hydrogen in its ground state 
\citep[see Eq.~(5) and Fig.~2 of][]{nussb+89}. 
The attenuation of the far-UV spectrum by Rayleigh scattering
thus provides an independent estimate of $N_{\rm H}$. 
It measures exclusively the amount of neutral hydrogen 
within the ISM. From this point of view, $N_{\rm H}$ derived 
from fitting the X-ray data can be larger by the CSM component 
than that obtained from fitting the Rayleigh attenuated 
continuum (cf. Eq.~(\ref{eq:nh})). 
In the case, when a SSS does not produce any nebular emission 
(e.g. RX\,J0439.8-6809), $N_{\rm H}$ values from both 
approaches should be equal. 

For $N_{\rm H}$ of a few $\times 10^{20}$\cmd, the effect of 
the Rayleigh attenuation creates an absorption core with 
the zero rest intensity and FWHM\,$\sim 10$\,\AA\ around 
the Ly-$\alpha$ line. It is thus 
well measurable on medium/high resolution spectra 
(e.g. by the HST/GHRS and FUSE spectrograph), but hardly 
detectable on the low resolution IUE spectra, because of 
influence of other absorption effects and the presence of 
the geocoronal Ly-$\alpha$ component. Here, an example is 
shown for RX\,J0439.8-6809 
(Fig.~\ref{fig:rx0439ray}, also Paper~II). 

\subsection{The nebular continuum}

The nebular component of radiation in the UV/optical continuum 
can be approximated by processes of recombination and thermal 
bremsstrahlung in the circumstellar material ionized by the 
SSS. In this case the energy distribution in the nebular 
continuum is determined by the total volume emission coefficient 
$\varepsilon_{\lambda}(T_{\rm e})$ 
($\rm\,erg\,cm^{3}\,s^{-1}\,\AA^{-1}$), i.e. the sum of 
contributions from recombination and bremsstrahlung for 
considered ions (mostly H$^{+}$, He$^{+}$ and He$^{++}$). 
Its profile depends on the electron temperature, $T_{\rm e}$, 
and is a function of the wavelength \citep[e.g.][]{b+m70}. 
Thus, the observed nebular flux can be expressed as 
a simple scaling of the emission coefficient, i.e., 
%
%
\begin{equation}
 F^{\rm obs}_{\rm n}(\lambda) = k_{\rm n} \times
      \varepsilon_{\lambda}(T_{\rm e}). 
\label{eq:fneb}
\end{equation}
%
According to the meaning of the
$\varepsilon_{\lambda}(T_{\rm e})$ coefficient, and 
assuming that $T_{\rm e}$ and thus 
$\varepsilon_{\lambda}(T_{\rm e})$ 
are constant throughout the nebula, the flux produced by 
the optically thin nebula 
(i.e. the second term in Eq.~(\ref{eq:fsss})), 
$F_{\rm n}(\lambda) = 
4\pi d^2 F^{\rm obs}_{\rm n}(\lambda) = 
\varepsilon_{\lambda}\int_V n_{\rm e}n_{+}{\rm d}V =
\varepsilon_{\lambda}$\textsl{EM}, where \textsl{EM} is
the so-called emission measure. Using its definition, 
Eq.~(\ref{eq:fneb}) can be expressed as 
%
%
\begin{equation}
     F^{\rm obs}_{\rm n}(\lambda) = \frac{EM}{4\pi d^2} \times
               \varepsilon_{\lambda}(T_{\rm e}),
\label{eq:fneb2}
\end{equation}
%
which determines the fitting parameter $k_{\rm n}$ as 
the emission measure scaled by the distance. Further fitting 
parameter is the electron temperature $T_{\rm e}$. 
Selection of an appropriate emission coefficient 
$\varepsilon_{\lambda}(T_{\rm e})$ depends on the properties 
of the nebula. To get the first estimate of the nebular 
parameters, contributions only from hydrogen plasma can be 
considered. Also it is difficult to separate contributions 
from the He\,\I\ continuum on the \textsl{IUE} and 
\textsl{HST} spectra, because of a very similar profile 
of its recombination coefficients to that of H\,\I\ 
\citep[e.g.][]{b+m70} and a small abundance. In the case of 
recognizable signatures of the He\,\I\I\ continuum 
(e.g. a jump in emission at $\sim$2050\,\AA\ and a pronounced 
Paschen series of the He\I\I\ recombination lines), the 
emission coefficient 
$\varepsilon_{\lambda}({\rm H, He^{+}},T_{\rm e})$
should be used, with the possibility to estimate the abundance 
of He$^{++}$ ions. Finally, it is reasonable to assume that 
the nebula is sufficiently dense ($n \gg 10^3$\cmt), which 
excludes contributions due to the two-photon emission 
\citep[e.g.][]{gur}. 
%
%
%

\subsection{The giant contribution}

If a SSS contains an M-type giant, its contribution usually 
rivals that from the nebula in the optical, but dominates 
the near-IR \citep[see Figs.~2-22 of][]{sk05}. Therefore, it 
is satisfactory to figure the observed SED of the giant 
by the broad-band $(VRI)JHKLM$ photometric fluxes. With 
respect to a large difference in the SED between a simple 
blackbody and the true giant's photosphere, it is suggested 
to approximate the observed fluxes from the giant, 
$F_{\rm g}^{\rm obs}(\lambda)$, with an appropriate synthetic 
spectrum, $\mathcal{F}_{\lambda}^{\rm synth.}(T_{\rm eff})$. 
Then we can write 
%
%
\begin{equation}
 F_{\rm g}^{\rm obs}(\lambda) = 
       \mathcal{F}_{\lambda}^{\rm synth.}(T_{\rm eff}) = 
        \theta_{\rm g}^2 F_{\rm g}(\lambda), 
\label{eq:fgiant}
\end{equation}
%
where the angular radius of the giant $\theta_{\rm g}$
(= $R_{\rm g}/d$) scales its emitted spectrum $F_{\rm g}(\lambda)$ 
to that observed at the Earth. The fitting parameters are 
the effective temperature $T_{\rm eff}$ and $\theta_{\rm g}$. 
These define the observed bolometric flux, 
$F_{\rm g}^{\rm obs} = \theta_{\rm g}^2 \sigma T_{\rm eff}^4$ 
\citep[see Eq.~(3) of][]{sk05} and thus the luminosity of 
the giant, $L_{\rm g} = 4\pi\ d^2 F_{\rm g}^{\rm obs}$. 

\subsection{The SED-fitting analysis}

Having defined individual components of radiation in the 
composite spectrum (Eq.~(\ref{eq:fsss})), we can model the 
X-ray---IR fluxes, as observed at the Earth, with a function 
%
%
\begin{equation}
 F^{\rm mod}(\lambda) = 
             F_{\rm h}^{\rm obs}(\lambda) + 
             F_{\rm n}^{\rm obs}(\lambda) +
             F_{\rm g}^{\rm obs}(\lambda), 
\label{eq:fmod}
\end{equation}
%
%
whose terms are defined by Eqs.~(\ref{eq:fhot}), (\ref{eq:fneb}) 
and (\ref{eq:fgiant}). The fitted fluxes consist of the 
{\em absorbed} X-ray fluxes (Eq.~(\ref{eq:ftaux})) and 
{\em dereddened} UV/IR fluxes (Eq.~(\ref{eq:extin})). 
If applicable, it is convenient to subtract the observed fluxes 
from the giant, which simplifies the task. To find the best 
solution, a grid of models SED (\ref{eq:fmod}) is calculated 
for reasonable ranges of the fitting parameters 
($\theta_{\rm h}$, $T_{\rm h}$, $N_{\rm H}$ for the SSS 
and $k_{\rm n}$, $T_{\rm e}$ for the nebula) to select that 
corresponding to the minimum of the function 
%
%
\begin{equation}
 \chi^{2} = \sum_{\rm i=1}^{\rm N}
    \left[\frac{(F^{\rm obs}(\lambda_{\rm i}) -
            F^{\rm mod}(\lambda_{\rm i})}
            {\Delta F^{\rm obs}(\lambda_{\rm i})}\right]^{2},
\label{eq:chi2}
\end{equation}
%
where $F^{\rm obs}(\lambda_{\rm i})$ are observed fluxes of 
the continuum (i.e. the absorbed X-ray fluxes and dereddened 
UV/IR fluxes), $N$ is their number, 
$\Delta F^{\rm obs}(\lambda_{\rm i})$ are their errors and 
$F^{\rm mod}(\lambda_{\rm i})$ are the theoretical fluxes 
given by Eq.~(\ref{eq:fmod}). 

In the following section I demonstrate this multiwavelength 
SED-fitting analysis on four selected objects. Their list with 
the log of the used spectroscopic observations are found in 
Table~1. In all cases, typical values for flux deviations of 
$\sim 10$\% were adopted. Resulting parameters are introduced 
in Table~2 and the corresponding models are depicted in 
Figs.~1,2,4 and 5. 

\section{Application to selected SSS}

\subsection{The SMC symbiotic X-ray binary RX\,J0059.1-7505}

Originally, this star was included in ``A New Catalogue of 
Emission-Line Stars and Planetary Nebulae in the Small 
Magellanic Cloud'' by \cite{lindsay} under the number 358. 
According to 
\cite{san+pesch81}, LIN~358 has a late K or early M-type 
continuum, displaying a variable \ha\ emission. Later, on 
the basis of the optical observations, \cite{walker83} found 
that the LIN~358 spectrum satisfies characteristics of a symbiotic 
star. \cite{nuss+vog95} analyzed the \textsl{HST FOS} spectra 
of LIN~358 and estimated its temperature within the range 
of $1.50 - 1.75 \times 10^5$\,K and the luminosity between 
2\,400 and 3\,600\lo. 
\cite{muerset+96}, analyzing ultraviolet and optical spectroscopy 
and infrared photometry, derived the spectral type of the giant 
as mid K, its temperature to 4000\,K, the luminosity of 
5\,500\,\lo\ and the radius of 150\ro. 
Using the Zanstra method, they estimated 
 $T_{\rm h} = 1.4\times 10^{5}$\,K, 
 $L_{\rm h} = 1.5\times 10^{37}$\es\ 
and 
 $R_{\rm h} = 0.11$\ro\ for the hot component. 
\cite{muerset+97} found that the pointed \textsl{ROSAT PSPC} 
observations of RX~J0059.1-7505 coincides in the position 
with the symbiotic star LIN~358. They classified its X-ray 
spectrum as an $\alpha$-type, i.e. with supersoft distribution. 
RX~J0059.1-7505 is presented in the catalogue of 
\cite{haberl+00} as a SSS. 

\cite{kah+hab06} observed RX\,J0059.1-7505 with \textsl{XMM-Newton} 
on 2003 November 16/17. Their blackbody fit to the supersoft 
component (0.13 -- 1.0 keV) corresponded to
the hot source temperature, 
 $T_{\rm h} = 227\,500\pm 30000$\,K, 
its radius, 
 $R_{\rm h} = 0.23^{+0.08}_{-0.04}$\ro\ 
and luminosity, 
 $L_{\rm h} = 1.0\times 10^{38}$\es\ for 
the total absorption parameterized with 
 $N_{\rm H} = 7.6\times 10^{20}$\cmd. 
The authors tried to extrapolate their X-ray flux distribution 
into the UV with the aim to compare their model to 
the \textsl{HST} observations (see their Fig.~4). 
The same observations were analyzed also by \cite{orio+07}, 
who obtained a blackbody model parameters, 
 $T_{\rm h} = 232000 $\,K, 
 $R_{\rm h} = 0.16$\ro, 
 $L_{\rm h} = 2.5\times 10^{38}$\es, 
and
 $N_{\rm H} = 7.0\times 10^{20}$\cmd. 
Their best fit with an atmospheric model corresponded to 
unrealistically high luminosity, 
 $L_{\rm X} \sim 8.3\times 10^{38}$\es\
within the (0.15 -- 1.0)\,keV range only 
($T_{\rm h} = 200000 $\,K, and
$N_{\rm H} = 8.9\times 10^{20}$\cmd). 
Applying the photoionization code \textsl{CLOUDY} to 
optical and ultraviolet \textsl{IUE} spectra, they 
estimated 
 $R_{\rm h} = 0.127$\ro, 
 $T_{\rm h} = 180000$\,K,
which corresponds to 
 $L_{\rm h} = 5.8\times 10^{37}$\es. 
Their parameters of the nebular component of radiation implied 
the emission measure, $EM = 1.3\times 10^{61}$\cmt. 
%
%
\begin{table}[p!t]
\caption[]{Log of the used spectroscopic observations}
\begin{center}
\begin{tabular}{ccccc}
\hline
\hline
    Date   & Julian date  & Region  & Observatory & $T_{\rm exp}$  \\
yyyy/mm/dd & JD~2\,4...   &  [nm]   &             &      [ks]       \\
\hline
\multicolumn{5}{c}{RXJ0059.1-7505 (LIN~358)}                       \\
2009/09/25 & 55099.5 & 3.1--8.4 &\textsl{XMM-Newton}$^{1}$  & 32 \\
2002/07/25 & 52480.5 & 92--119  &\textsl{FUSE}              & 26.4\\
1995/03/20 & 49796.5 &150--540  &\textsl{HST}               & 2.5 \\  
\multicolumn{5}{c}{RXJ0438.8-6809}                           \\
1992/01/03 & 48626.5 & 1.7--6.9 &\textsl{ROSAT}$^{2}$   & 2.7 \\
1998/11/17 & 51135.5 & 115--316 &\textsl{HST}           &2.1 \\
1995/10/15 & 50006.5 & 380--695 & La Silla, 3.6-m$^{3}$ &10.8 \\ 
\multicolumn{5}{c}{RX\,J2030.5+5237 (V1974~Cyg)} \\
1992/12/07 & 48964.5 & 1.24--4.13 & \textsl{ROSAT}$^{4}$ & 2.9 \\
1993/07/14 & 49183.5 & 1.24--4.13 & \textsl{ROSAT}$^{4}$ & 2.3 \\
1993/04/04 & 49082.3 & 115--198   & \textsl{IUE}   & 0.87\\
1993/07/23 & 49192.4 & 190--335   & \textsl{IUE}   & 1.5 \\
1993/07/24 & 49192.5 & 115--198   & \textsl{IUE}   & 1.5 \\
\multicolumn{5}{c}{RX\,J1601.6+6648 (AG~Dra)} \\
1993/04/15 & 49093.5  & 3.5--11.5 &\textsl{ROSAT}$^{5}$ & 2.5 \\
2004/06/15 & 53172.0  & 3.1--6.9  &\textsl{XMM-Newton}$^{6}$ &12.4 \\
2004/06/15 & 53172.0  & 212--291  &\textsl{XMM-OM}$^{6}$  &~4.4 \\
2005/06/11 & 53533.5  & 2.8--6.9  &\textsl{XMM-Newton}$^{6}$ & 18.3 \\
2005/06/11 & 53533.5  &   231     &\textsl{XMM-OM}$^{6}$ & 14.2 \\
2004/06/24 & 53181.0  & 100--108  &\textsl{FUSE}        &10.8 \\
1979/09/25 & 44143.0  & 116--325  &\textsl{IUE}         &~3.4 \\
1993/04/09 & 49087.4  & 116--325  &\textsl{IUE}         &0.48 \\
\hline
\end{tabular}
\end{center}
  $^{1}$~\cite{kah+hab06}, 
  $^{2}$~\cite{teeseling+99}, 
  $^{3}$~\cite{teeseling+96}, 
  $^{4}$~\cite{krautter+96}, 
  $^{5}$~\cite{greiner+97}, 
  $^{6}$~\cite{gon+08}
\end{table}
%
%
%
\begin{figure*}
\centering
\begin{center}
\resizebox{16cm}{!}{\includegraphics[angle=-90]{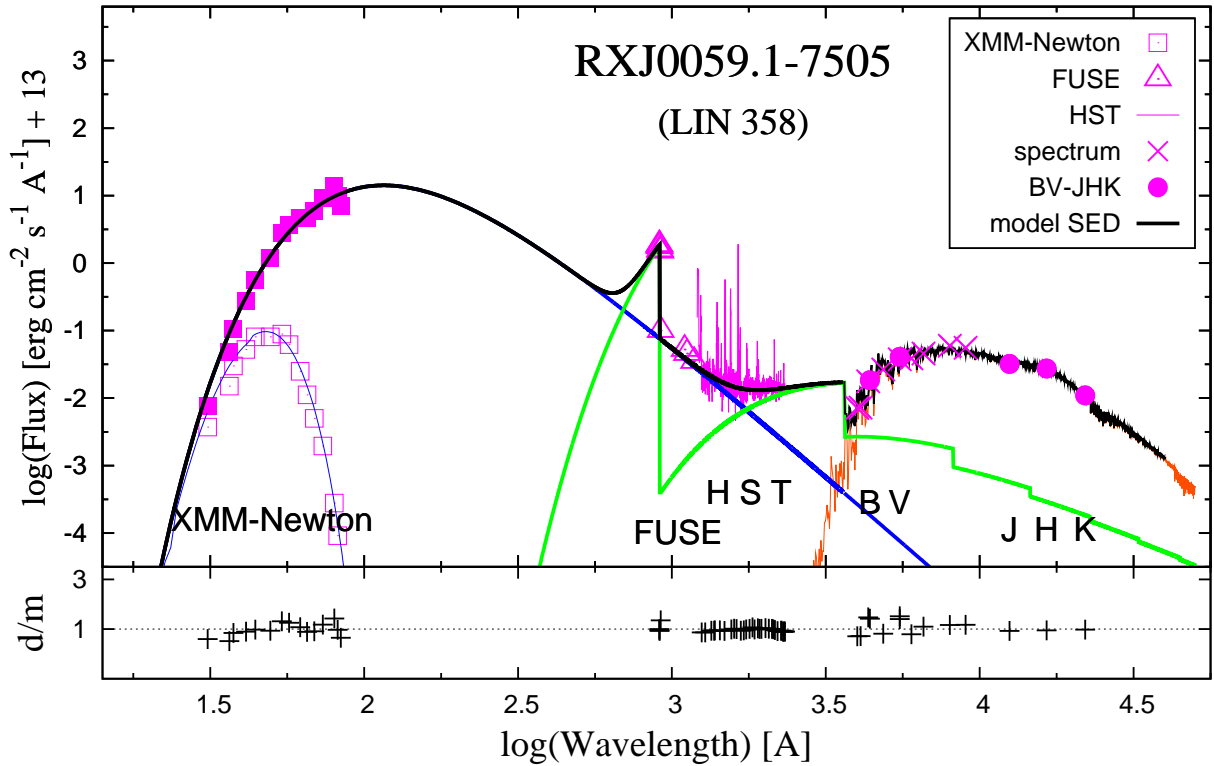}}
\end{center}
\caption[]{
 A comparison of the measured (in violet) and modelled 
 (heavy black line) SED of RX~J0059.1-7505 (LIN~358) 
 with corresponding data-to-model ratios (d/m). 
 Open and filled squares are the measured and unabsorbed X-ray 
 fluxes, respectively. 
 The model SED and its components of radiation here represent 
 a graphic form of Eq.~(\ref{eq:fmod}). The blue, green and 
 orange lines denote components of radiation from the SSS, 
 nebula and giant, respectively. 
          }
\label{fig:lin358sed}
\end{figure*} 

\subsubsection{Multiwavelength model SED of RX\,J0059.1-7505}

Spectroscopic observations used to model the SED of 
RX\,J0059.1-7505 cover the spectral range from the supersoft 
X-rays to the near-IR ($\sim 3.1-2200$\,nm, Table~1). 
The supersoft X-ray fluxes, taken with \textsl{XMM-Newton}, 
were reconstructed from Fig.~4 of \cite{kah+hab06}. 
The ultraviolet spectra with \textsl{FUSE} (E9500501), 
\textsl{HST FOS} (Y2CQ0203T, Y2CQ0204T, Y2CQ0205T) and 
\textsl{HST COS} (LAAZ01010) were obtained from the satellite 
archives with the aid of the Multimission Archive at the Space 
Telescope Science Institute (MAST). 
These observations were supplemented with the photometric
$BVJHK$ measurements published by \cite{muerset+96}. 
UV to IR data were dereddened with $E_{\rm B-V} = 0.08$\,mag 
and the corresponding parameters were scaled to the distance 
of SMC, 60\kpc\ \citep[][]{mateo98}. 

It was possible to match the photometric $BVJHK$ flux-points 
with a synthetic spectrum calculated for 
$T_{\rm eff} = 4000 \pm 200$\,K 
\citep[selected from a grid of models made by][]{h+99}, 
and scaled with $\theta_{\rm g} = 6.7\times 10^{-11}$, 
which corresponds to the bolometric flux, 
  $F_{\rm g}^{\rm obs} = 6.5\times 10^{-11}$\,\ecs, 
and thus the luminosity 
  $L_{\rm g} = 7\,300\,(d/60\,\kpc)^2$\lo, 
(i.e. absolute bolometric magnitude 
  $M_{\rm g}^{\rm bol}= -4.92$). 
The value of $\theta_{\rm g}$ gives the radius of the giant, 
  $R_{\rm g} = 178\,(d/60\,\kpc)$\ro. 
These parameters allow us to classify the cool giant in 
RX\,J0059.1-7505 as a K5\,Ib supergiant. 

After subtracting the contribution from the giant, I modelled 
15 representative supersoft X-ray fluxes between 31 and 84\,\AA\ 
and 30 ultraviolet continuum flux-points between 
1\,250 and 2\,330\,\AA\ by the function 
$F^{\rm mod}(\lambda) = 
             F_{\rm h}^{\rm obs}(\lambda) +
             F_{\rm n}^{\rm obs}(\lambda)$
(see Eq.~(\ref{eq:fmod})) 
to determine its variables 
 $\theta_{\rm h}$, 
 $T_{\rm h}$, $N_{\rm H}$, 
 $k_{\rm n}$ and $T_{\rm e}$. 
%
The best model has the reduced $\chi^{2} = 1.3$ 
(for 40 degrees of freedom), and is determined by the most 
probable fitting parameters, 
 $\theta_{\rm h} = 3.4\times 10^{-14}$, 
 $T_{\rm h} = 250000$\,K, 
 $N_{\rm H} = 6.1\times 10^{20}$\cmd, 
 $k_{\rm n} = 5.6\times 10^{12}$\,cm$^{-5}$ and 
 $T_{\rm e} = 18000$\,K, 
which yield 
 $R_{\rm h}^{\rm eff} = 0.09\,(d/60\,\kpc)$\ro, 
 $L_{\rm h} = 1.1\times 10^{38}\,(d/60\,\kpc)^2$\es 
and 
 $EM = 2.4\times 10^{60}\,(d/60\,\kpc)^2$\cmt. 
%

Observations with \textsl{FUSE} were not used directly in 
the fitting procedure, because of large uncertainties 
of the faint continuum for different exposures. Therefore, 
I compared only a few selected mean fluxes from LiF2A 
(1087 -- 1181\,\AA) channel spectrum to demonstrate the steep 
slope of the far-UV spectrum, where the SSS has a dominant 
contribution. 
Another very interesting feature of the \textsl{FUSE} spectrum 
is a strong Lyman jump in emission, which signals that the nebula 
is optically thin in the Lyman continuum. 
Its value is very sensitive to the electron temperature, which 
constrains its uncertainty only to $\pm 2000$\,K. However, the 
extinction curve is not well determined at this region 
(see \cite{c+89} and Sect.~2.2 of \cite{sk+06}). Therefore, 
I dereddened the data around 912\,\AA\ by multiplying 
them with a factor of 2.5, which places corrected fluxes 
for $\lambda > 912$\,\AA\ at the hot star model. 
Here I used the SiC1B channel spectrum (915 -- 992\,\AA). 

\subsection{The LMC X-ray source RX\,J0439.8-6809}

RX\,J0439.8-6809 is a bright SSS in the LMC, which was 
discovered in the \textsl{ROSAT} All-Sky-Survey during 
1990 November 16--29. Following pointing observations, 
performed on 1992 January 3-rd, allowed its more detailed 
study \citep[][]{greiner+94}. Fitting the combined survey 
and pointing data with a blackbody spectrum, the authors 
derived an effective temperature of the source, 
$kT_{\rm bb} = (20 \pm 10)$\,eV ($232000 \pm 115000$\,K), 
absorbed with $N_{\rm H} = (4.2 \pm 2)\times 10^{20}$\cmd, 
which, as they noted, is in excellent agreement with the 
galactic column density towards RX\,J0439.8-6809 of 
$4.5\times 10^{20}$\cmd. 
\cite{greiner+94} also found that RX\,J0439.8-6809 was 
remarkably constant during its monitoring with \textsl{ROSAT} 
($\sim 14$ months), exhibiting only a few days X-ray 
variability. 
\cite{teeseling+96} identified RX\,J0439.8-6809 with a very 
blue and faint (B = 21.5) object. They revealed that the steep 
profile of their optical spectrum corresponds to the 
Rayleigh-Jeans tail of the SSS component. Taking 
$N_{\rm H} = 4\times 10^{20}$\cmd\ they obtained acceptable 
blackbody fit to both the X-ray and the optical data 
with $T_{\rm bb} \approx 300000$\,K, a radius 
$R = 4.8\times 10^9$\,cm and a luminosity 
$L = 1.4 \times 10^{38}$\es. An additional interesting feature 
of the optical spectrum was the absence of any absorption 
or emission line. \cite{schmidtke+96} confirmed the optical 
counterpart to the X-ray source by the $UBV$ photometry, 
and revealed a low-amplitude 0.1403 and 0.1637-day period 
in the V-band. 
\cite{teeseling+97} investigated the nature of RX\,J0439.8-6809 
and concluded that it is most probably an accreting 
double-degenerate binary with an orbital period of a few 
minutes. 
Based on the \textsl{HST/STIS} ultraviolet spectroscopy, 
\cite{teeseling+99} estimated the neutral hydrogen column density 
to $(4\pm 1)\times 10^{20}$\cmd\ by fitting the broad Ly-$\alpha$ 
profile with a pure damping effect. They also performed 
a blackbody fit to the X-ray, UV and optical fluxes, 
corresponding to $T_{\rm bb} = 295000$\,K, 
$R = 5\times 10^9$\,cm and $L = 1.6 \times 10^{38}$\es\ 
for the distance of 50\kpc. 
Application of a more sophisticated NLTE model to the same 
observations, also yielded acceptable fits with similar 
parameters. 
Finally, a very good fit with a pure CO model 
($\log(g) = 7$, 
$T_{\rm eff} \sim 310000$\,K, 
$L \sim 3 \times 10^{38}$\es), 
the absence of long-term variability and the proximity of 
RX\,J0439.8-6809 to the theoretical carbon-burning main 
sequence, led the authors to speculate that RX\,J0439.8-6809 
represents a completely new type of star. 
%
%
\begin{figure}
\centering
\begin{center}
\resizebox{\hsize}{!}{\includegraphics[angle=-90]{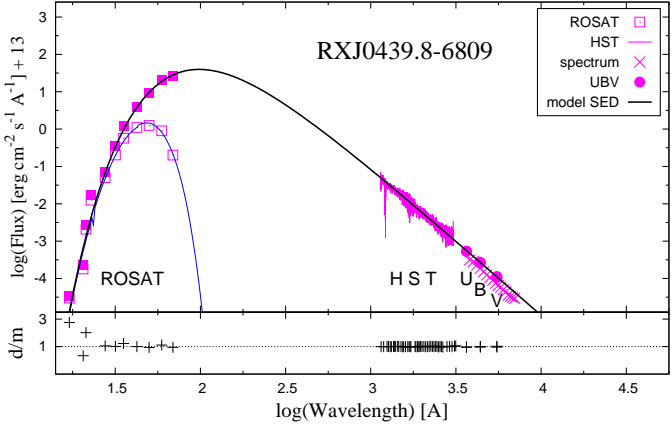}}
\end{center}
\caption[]{
As in Fig.~\ref{fig:lin358sed}, but for RX\,J0439.8-6809. 
The \textsl{ROSAT}, \textsl{HST} and optical fluxes could 
be fitted simply only with the SSS component 
($F_{\rm h}^{\rm obs}(\lambda)$ in Eq.~(\ref{eq:fmod})). 
          }
\label{fig:rx0439sed}
\end{figure} 

\subsubsection{Multiwavelength model SED of RX\,J0439.8-6809}

Observations used to model the SED of RX\,J0439.8-6809 cover 
the spectral range from the supersoft X-rays to the optical 
\ubv flux-points ($\sim 1.7-5500$\,nm). 
The supersoft X-ray fluxes were 
taken from Fig.~5 of \cite{teeseling+99}. The ultraviolet 
\textsl{HST/STIS} spectrum (O55G01010) was obtained from 
the satellite archive with the aid of the MAST. 
Spectroscopic observations were supplemented with the 
\ubv broad-band photometry according to \cite{schmidtke+96} 
and the catalog of \cite{zaritski} 
(U = 20.068, B = 21.361, V = 21.534). 
Ultraviolet and optical data were corrected for interstellar 
extinction with $E_{\rm B-V} = 0.06$\,mag and the 
corresponding parameters were scaled to the distance 
of LMC, 49\kpc\ \citep[][]{mateo98}. 

Fitting a simple blackbody radiation to 61 selected fluxes 
(11 X-ray, 44 UV and 6 \ubv fluxes) with the same weight, 
yielded 
 $T_{\rm h} = 295000 \pm 5000$\,K, 
 $N_{\rm H} = (4.2 \pm 0.2) \times 10^{20}$\cmd\
and 
 $\theta_{\rm h} = (3.7\pm 0.2)\times 10^{-14}$, 
which corresponds to 
 $R_{\rm h}^{\rm eff} = 0.08\pm 0.01 (d/49\,\kpc)$\ro\ 
and 
 $L_{\rm h} = (1.7\pm 0.2)\times 10^{38}(d/49\,\kpc)^2$\es. 
The resulting model SED with the used observations are shown 
in Fig.~\ref{fig:rx0439sed}. The Planck curve matches 
the selected unabsorbed and dereddened fluxes with the 
reduced $\chi^2 = 2.9$ for 10\% errors adopted for all fluxes, 
but with only $\chi^2 = 0.96$, when enlarging errors to 
20--30\% in the first 3 fluxes at 17.0, 20.5 and 21.3\,\AA\ 
\citep[see Fig.~5 of][]{teeseling+99}. 
The optical spectrum of \cite{teeseling+96} was not included 
in the fitting, because its continuum was a factor of 
$\sim$1.2 below the \ubv fluxes that follow exactly the model 
of the UV spectrum. However, the steep slope of the optical 
spectrum is very similar to that given by the \ubv flux-points 
(see Fig.~\ref{fig:rx0439sed}). 
%
%
\begin{figure}
\centering
\begin{center}
\resizebox{\hsize}{!}{\includegraphics[angle=-90]{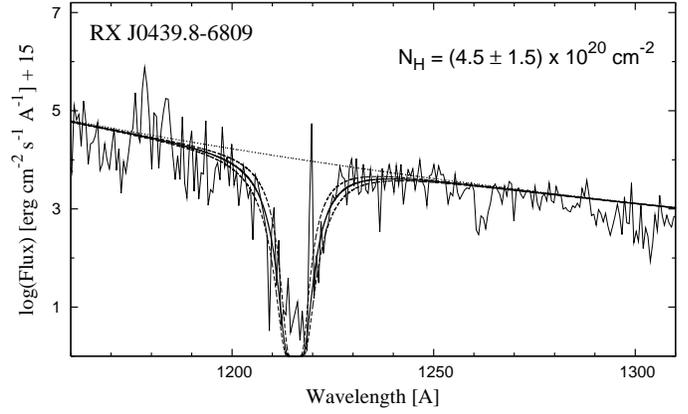}}
\end{center}
\caption[]{
Dereddened \textsl{HST} spectrum of RX\,J0439.8-6809 (the gray 
line). The solid thick line and dashed lines represent the model 
and its deviations, respectively, of the absorption feature 
around the Ly-$\alpha$ line due to the Rayleigh scattering 
(see Sect.~2.3.1). The dotted line is the non-scattered blackbody 
radiation from Fig.~\ref{fig:rx0439sed}. 
          }
\label{fig:rx0439ray}
\end{figure}
%

According to the suggestion in Sect.~2.3.1., I matched 
the attenuated continuum around the Ly-$\alpha$ line 
with the Rayleigh scattering process. The fit corresponds 
to $N_{\rm H} = (4.5 \pm 1.5) \times 10^{20}$\cmd\ 
(see Fig.~\ref{fig:rx0439ray}), which is equal to that 
derived independently from modelling the SED. This suggests 
that there is no CSM absorption component in the direction of 
RX\,J0439.8-6809, because the Rayleigh scattering measures only 
the ISM component. This result is consistent with the fact that 
there is no nebular component of radiation indicated in the 
spectrum of this SSS. 

Modelling the SED of RX\,J0439.8-6809 represents the simplest 
case here presented. Its radiation dominates the entire 
observed spectrum (1.7--680\,nm), which allows to determine 
unambiguously all the fitting parameters with relatively small
uncertainties (Table~2). 
%
%
\begin{figure*}
\centering
\begin{center}
\resizebox{16cm}{!}{\includegraphics[angle=-90]{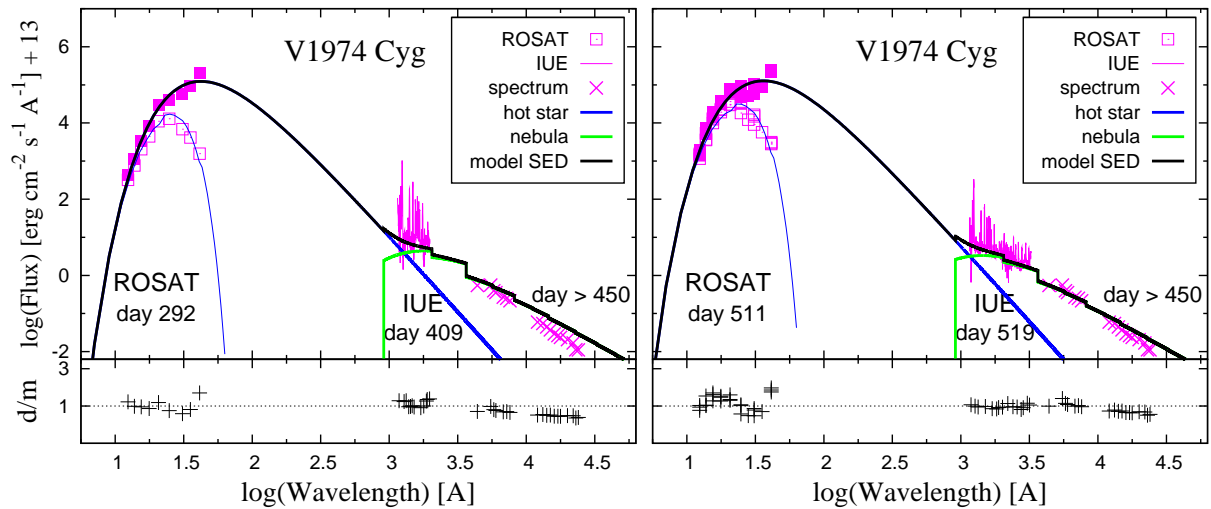}}
%
\end{center}
\caption[]{
Example of the multiwavelength modelling the soft X-ray---near-IR 
SED of the classical nova V1974~Cyg during its SSS phase. 
          }
\label{fig:v1974sed}
\end{figure*} 

\subsection{Classical nova V1974~Cyg}

The classical nova V1974\,Cyg (Nova Cygni 1992, RX\,J2030.5+5237) 
was discovered by \cite{c92} on 1992 February 19.07 UT. It 
reached a peak visual magnitude of 4.5 on 1992 February 21.01 UT 
\citep[][]{sch92}. It was classified as a moderately fast 
O-Ne-Mg nova with $t_{\rm 3,V} = 42$ days 
\citep[][]{chochol+93,shore+93}. The distance to the nova can 
be put to 1.8\kpc\ \citep[][]{chochol+93,paresce+95} and the 
colour excess $E_{\rm B-V}$ = 0.36 \citep{austin+96}. 

V1974~Cyg was observed across the entire electromagnetic 
spectrum, from the $\gamma$-rays to the radio 
\citep[see][for a review]{austin+96}. Extensive observations 
in the optical and ultraviolet wavelengths showed that the 
nova entered the nebular phase from 1992 April ($\gtrsim 50$ days 
after the optical maximum), and from 1992 September 
($\gtrsim 200$ days) the nebular lines dominated its spectrum 
\citep[e.g.][]{shore+93,chochol+93,barger+93}. 

Observations in the X-ray domain were carried out with the 
\textsl{ROSAT} satellite from 1992 April 22 to 1993 December 3. 
\cite{krautter+96} analyzed first the X-ray observations. They 
found that a soft component ($\sim 0.1 - 1.0$\,keV) dominated 
the emitted energy during 255--511 days after the optical 
maximum (see their Fig.~1), and had all the characteristics 
of a SSS. Their blackbody models fitted well the measured SED 
of the soft X-ray component in the range of 0.2--1\,keV. 
However, their best fits gave column densities of the 
order of several times $10^{21}$\cmd, effective temperatures 
$kT = 19 - 26$\,eV (i.e. 220600 -- 302000\,K), but totally 
unrealistic bolometric luminosities of several thousand of 
$L_{\rm Edd}$ for a 1\mo\ WD and the distance of 1.7\kpc. 
Therefore, the authors judged these parameters as unreliable 
and suggested that results obtained from blackbody fits 
to the supersoft X-ray sources should be used with care. 
To satisfy theoretical calculations they adopted the Eddington 
luminosity and the temperature of $3\times 10^{5}$\,K for 
the postnova WD in V1974~Cyg. 
The \textsl{ROSAT} spectra were reanalyzed by \cite{balman}, 
who compared them with hydrostatic LTE atmosphere models for 
a 1.2\mo\ WD. They determined its luminosity to be a factor of 
$\sim$2 below the Eddington one and its temperature to 590000\,K 
(day 511), assuming $N_{\rm H} = 2.0-2.2\times 10^{21}$\cmd. 

\subsubsection{Multiwavelength model SED of V1974~Cyg}

To demonstrate the multiwavelength modelling the SED for the 
classical nova V1974~Cyg, I reanalyzed its energy spectra made 
on day 292 and 511 presented in Fig.~4 of \cite{krautter+96}. 
As the SSS phase of V1974~Cyg lasted for a long time, from 
day $\sim$255 to $\sim$511 after the optical maximum, the X-ray 
data were complemented with observations in other wavelengths 
made during this period, being the nearest to day 292 and 
511, respectively. In particular, the day 292 was complemented 
with the well exposed \textsl{IUE} spectra SWP47416(7) and the 
day 511 with 
SWP48219(20) and LWP25981(2) spectra. Some optical/near-IR fluxes 
were extracted from the ground-based spectroscopic observations 
of \cite{rafanelli+95} from 1993 March 19 (day 393, 
$\lambda\lambda 580-760$\,nm) and of \cite{wagner+96} from 
1993 May 2 (day 438, $\lambda\lambda 1200-2400$\,nm). Finally, 
photometric $BV$ flux-points were derived from observations 
published by \cite{chochol+93}. 
Due to a rich and very strong emission line spectrum of the 
nova, the $BV$ magnitudes were corrected for emission lines to 
get fluxes of the real continuum \citep[see Table~2 in][]{sk07}. 
The total observed spectrum covers the range from 1.24\,nm to 
2400\,nm. It was fitted by the function, 
             $F^{\rm mod}(\lambda) = 
             F_{\rm h}^{\rm obs}(\lambda) + 
             F_{\rm n}^{\rm obs}(\lambda)$ 
(Eq.~(\ref{eq:fmod})). 
Fluxes at $\lambda > 760$\,nm suffer with a systematic 
deviation with respect to the predicted model, so their errors 
were formally enlarged to 30\%. 
The large value of the reduced $\chi^2$ results mainly 
from very uncertain measurements at the low energies of the 
X-ray spectrum \citep[][]{krautter+96}. 

The resulting parameters are given in Table~2 and the observed 
and model SED are depicted in Fig.~\ref{fig:v1974sed}. 
The results are discussed in Sect.~4. 
%
%
\begin{figure*}
\centering
\begin{center}
%
\resizebox{17cm}{!}{\includegraphics[angle=-90]{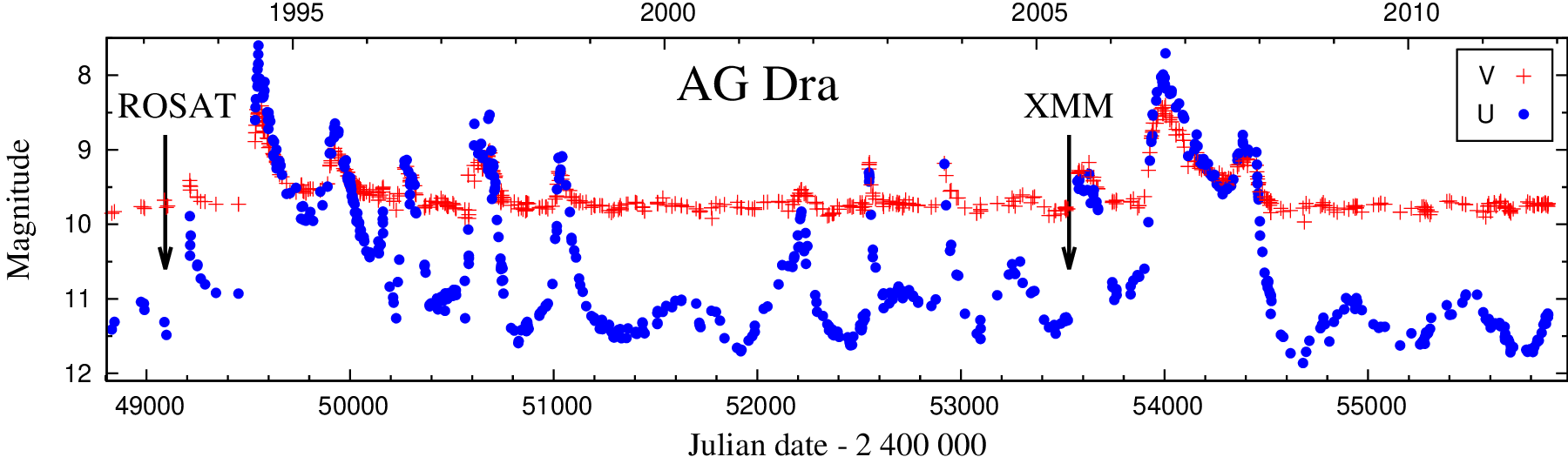}}\vspace{3mm}
\resizebox{17cm}{!}{\includegraphics[angle=-90]{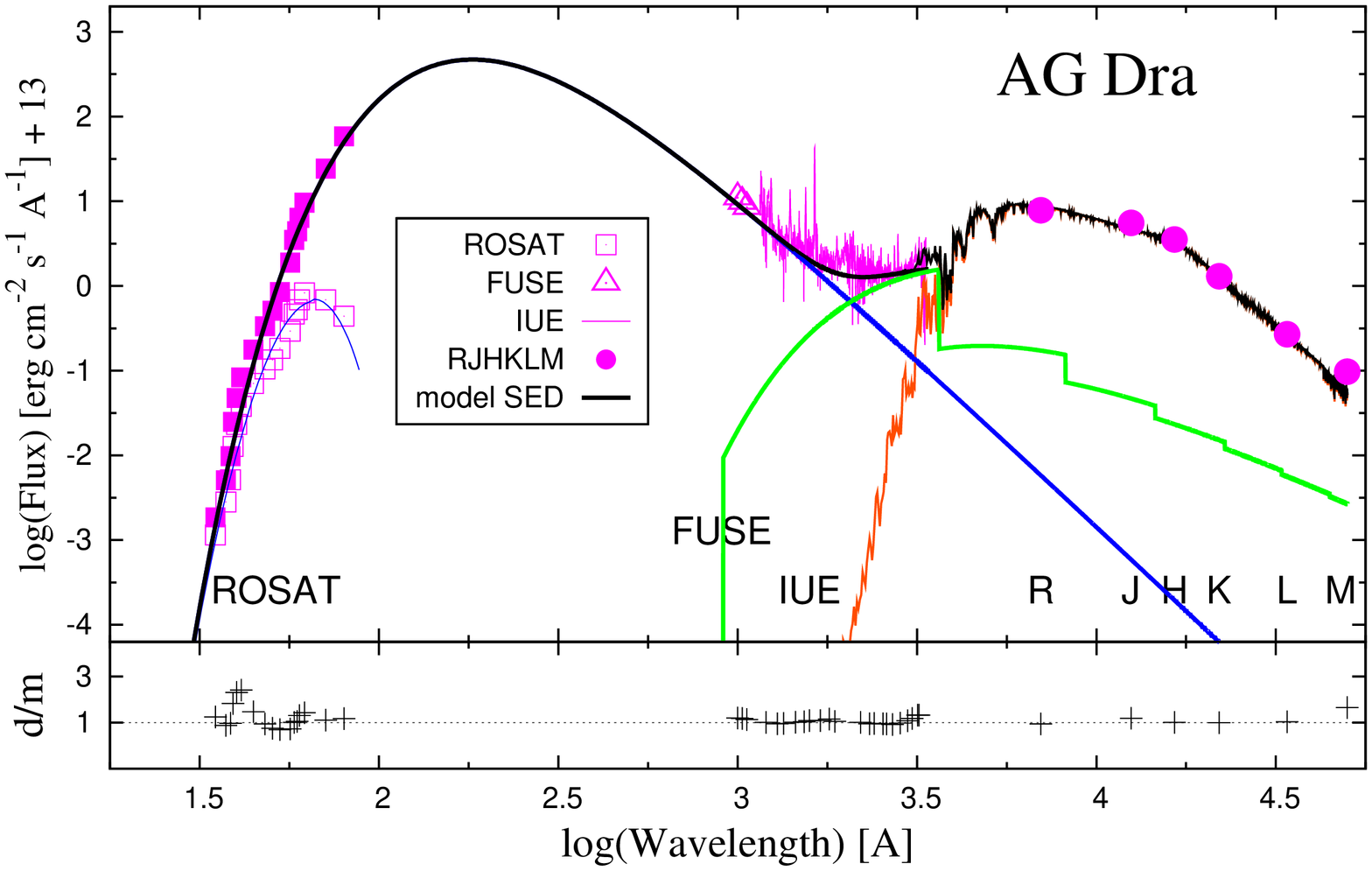}\hspace{5mm}
                      \includegraphics[angle=-90]{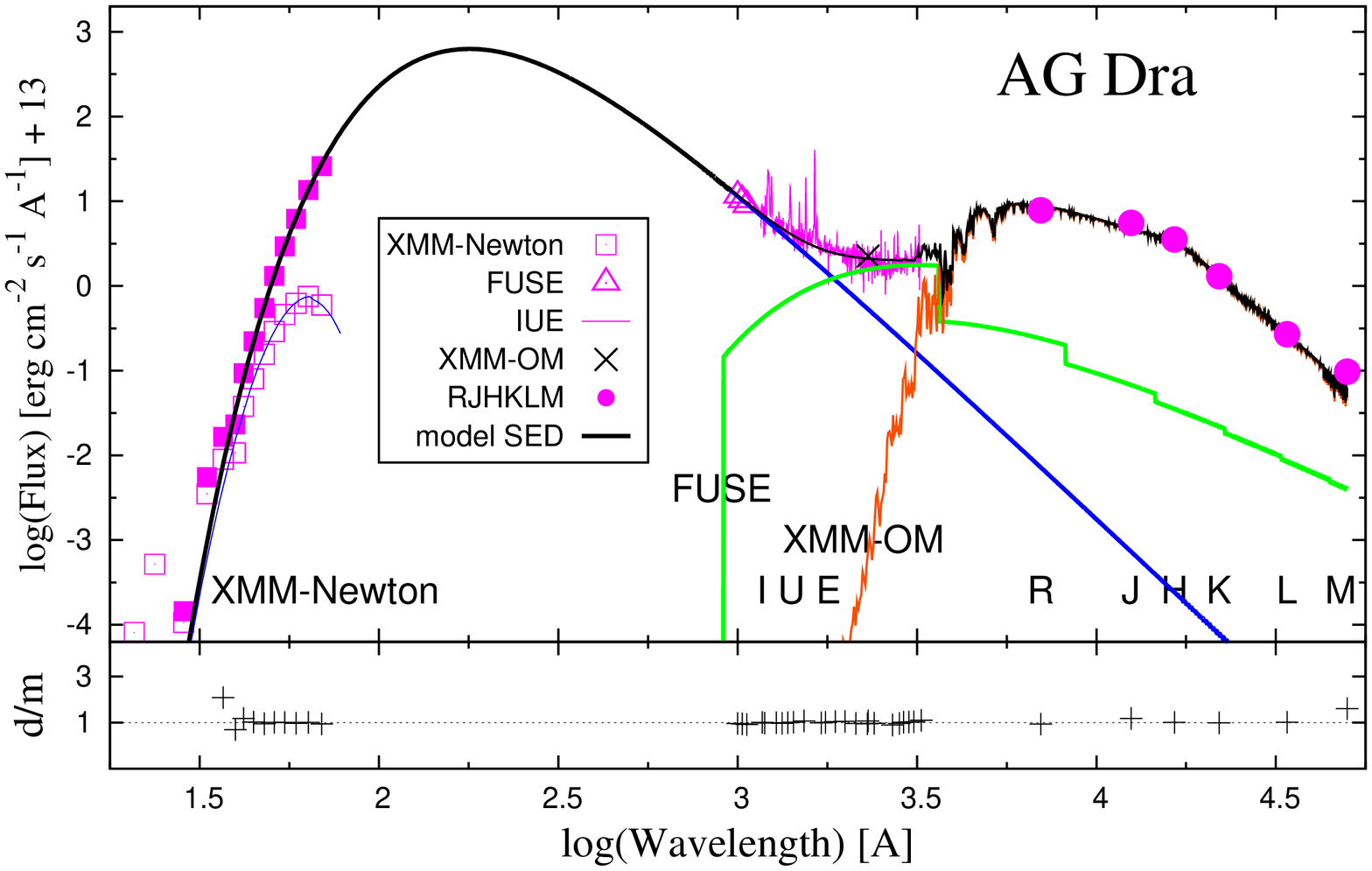}}
\end{center}
\caption[]{
Top panel shows the $U$ and $V$ light curves of AG~Dra from 1992.5. 
During quiescent phases ($U \gtrsim 11$), a strong supersoft X-ray 
emission from the burning WD is detected. The arrows mark the dates 
with the \textsl{ROSAT} and \textsl{XMM-Newton} observations, whose 
multiwavelength models SED are shown in bottom panels. Denotation 
of lines and observed fluxes is the same as in 
Fig.~\ref{fig:lin358sed}. 
          }
\label{fig:agdra}
\end{figure*}

\subsection{Classical symbiotic star AG~Dra}

AG~Dra (RX\,J1601.6+6648) is a classical symbiotic star. 
It is classified as a yellow symbiotic binary, because it 
comprises a K2\,III giant \citep{ms99} as the donor component. 
The accretor is a low mass WD accreting from the giant's wind 
on a 549-day orbit \citep[e.g.][]{f+00}. 
The optical light curve of AG~Dra shows numerous bursts with 
amplitude of 1--3\,mag in $U$, which are abandoned with large 
periods of quiescent phases 
\citep[e.g.][]{meinunger79,leedjarv+04,sk+12}. 
Modelling the UV/IR continuum, \cite{sk05} found a significant
contribution from the nebula in the near-UV/optical that 
strengthens during outbursts. 
%
AG~Dra is a halo binary system with a low orbital inclination 
\citep[][]{ss97,gon+08}, a low reddening 
\citep[$E_{\rm B-V} = 0.08\pm 0.01$\,mag,][]{birriel+00} 
and a low interstellar absorption 
\citep[$N_{\rm H} \sim 3\times 10^{20}$\cmd,][]{anderson+81}. 
These properties makes AG~Dra the brightest system in the 
supersoft X-ray energies among other galactic symbiotics. 

Therefore, AG~Dra has been frequently observed by the X-ray 
satellites, 
\citep[\textsl{Einstein}, \textsl{EXOSAT}, \textsl{ROSAT}, 
\textsl{XMM-Newton}, see ][]{greiner+97,gon+08}. 
\cite{greiner+97} modelled the \textsl{ROSAT}~PSPC observations 
during quiescent phase between 1992 April 16 to 1993 May 12. 
Using the blackbody model for fixed galactic value of 
$N_{\rm H} = 3.15\times 10^{20}$\cmd, they obtained 
$kT = 14.5$\,eV (168300\,K) and 
$L_{\rm h} = (9.5 \pm 1.5)\times 10^{36} (d/2.5\kpc)^2$\es. 
Analyzing the \textsl{ROSAT} data together with \textsl{IUE} 
and optical observations, they revealed an anticorrelation 
between the supersoft X-ray and the UV/optical fluxes. During 
the optical maxima (1994 and 1995) the X-ray flux decreased 
remarkably, while during the following quiescence it recovered 
to the pre-outburst value. 
The X-ray emission of AG~Dra was reviewed and discussed by 
\cite{gon+08}. They found that the anticorrelation between the 
X-ray and optical/UV emission appears to be a general feature
of AG~Dra radiation and is independent of the type of the 
outburst. They suggested that the WD radiation increases during 
outbursts, but is strongly absorbed by the circumstellar ionized 
gas. Modelling the X-ray---UV/optical continuum during different 
stages of the AG~Dra activity, \cite{sk+09} suggested that 
the flux anticorrelation is caused by the enhanced wind from 
the hot star as it is indicated for symbiotic binaries during 
active phases \citep[][]{sk06}. The wind absorbs the supersoft 
X-ray photons via the {\rm b--f} transitions, and via the 
{\rm f--b} and {\rm f--f} transitions enhances the nebular 
emission in the near-UV/optical domain. 
%

\subsubsection{Multiwavelength model SED of AG~Dra}

Two examples of the multi-band modelling the SED of AG~Dra 
during its quiescent phase, i.e. with a strong supersoft 
X-ray component, were selected to demonstrate the method 
on a classical symbiotic star. The first model is composed 
of the \textsl{ROSAT} PSCP observation \citep[][]{greiner+97} 
taken on 1993 April 15 and a nearly-simultaneous ultraviolet 
observation with \textsl{IUE} from 1993 April 9. 
The second one includes the supersoft X-ray data made with 
\textsl{XMM-Newton} and its optical monitor (\textsl{XMM-OM}) 
from 2005 June 11 \citep[][]{gon+08}, the far-UV observations 
with \textsl{FUSE} taken during quiescent phase on 2004 June 24 
and the \textsl{IUE} spectra, taken during quiescence at 
a similar orbital position (1979 September 25--27). In spite 
of a different time, their fluxes at the near-UV overlap those 
measured with \textsl{XMM-OM}. 
Both sets of observations were complemented with the 
$RJHKLM$ photometric flux-points, which define the cool giant 
radiation. Its model SED was adopted according to \cite{sk05}. 
Assuming that the radiation from the giant is constant, its 
contribution was subtracted from the near-UV fluxes, which 
reduced the model SED to 
$F^{\rm mod}(\lambda) = 
             F_{\rm h}^{\rm obs}(\lambda) + 
             F_{\rm n}^{\rm obs}(\lambda)$ 
(see Eq.~\ref{eq:fmod}). 
Resulting models and the corresponding parameters are depicted 
in Fig.~\ref{fig:agdra} and given in Table~2, respectively. 
They are scaled with the distance to AG~Dra of 1.1\,kpc 
\citep[][]{sk05}. Relatively constant values of the X-ray and 
the far-UV fluxes, measured at very different dates, reflect  
a stability of the WD's radiation in AG~Dra (Table~2). 
%
%
\begin{table*}[p!t]
\begin{center}
\caption{Physical parameters of selected SSSs derived by 
         the multiwavelength modelling their X-ray---IR 
         SED in the continuum. 
        }
\begin{tabular}{cc|ccc|cccc|cc|c}
\hline
\hline
 Object                         &
 Distance                       &
\multicolumn{3}{c|}{Giant}      &
\multicolumn{4}{c|}{Hot star (SSS)}   &
\multicolumn{2}{c|}{Nebula}     &
                                \\
                                &
$d$                             &
$R_{\rm g}$                     & 
$T_\mathrm{eff}$                & 
$L_{\rm g}$                     & 
$N_{\rm H}$                     & 
$R_{\rm h}^{\rm eff}$           & 
$T_{\rm h}$                     & 
$\log(L_{\rm h})$               & 
$T_{\rm e}$                     & 
$EM$                            & 
$\chi^2_{\rm red}$ / d.o.f.     \\
                                & 
[ kpc ]                         &
[ $R_{\odot}$ ]                 & 
[ K ]                           & 
[ $L_{\odot}$ ]                 &
[ $10^{20}\,{\rm cm^{-2}}$ ]    &
[ $R_{\odot}$ ]                 &
[ kK ]                          &
[ ${\rm erg\,s^{-1}}$ ]         &
[ kK ]                          & 
[ $10^{60}\,{\rm cm^{-3}}$ ]    & 
                                \\
%
%
\hline
LIN~358& 60 & 178$\pm 18$  & 4000$\pm 200$ & 7300$\pm 2000$ &
          6.1$\pm 0.2$ & 0.089$\pm 0.01$ & $250\pm 10$ & 38.03$\pm 0.11$ &
          18$\pm 5$ & 2.4$\pm 0.3$ & 1.3 / 40   \\
RX\,J0439& 49 & -- & -- & -- &
        4.2$\pm 0.2$ & 0.08$\pm 0.01$ & $295\pm 5$ & 38.23$\pm 0.05$ &
            -- & -- & 0.96 / 58   \\
V1974~Cyg$^{a})$& 1.8 & -- & -- & -- &
        18$\pm 4$ & 0.020$\pm 0.003$ & $685\pm 30$ & 38.49$\pm 0.15$ &
        40$\pm 10$ & 0.39$\pm 0.06$ & 15 / 17  \\
V1974~Cyg$^{b})$& 1.8 & -- & -- & -- &
        17$\pm 5$ & 0.014$\pm 0.002$ & $800\pm 30$ & 38.43$\pm 0.16$ &
        50$\pm 15$ & 0.30$\pm 0.05$ & 11 / 21  \\
AG~Dra$^{c})$& 1.1 &34$\pm 4$ &4300$\pm 200$ &360$\pm 90$ &
        2.90$\pm 0.08$ & 0.029$\pm 0.003$ & $159\pm 2$ & 36.28$\pm 0.10$ &
        15.0$\pm 2.0$ & 0.060$\pm 0.005$ & 5.4 / 39   \\
AG~Dra$^{d})$& 1.1 & \multicolumn{3}{c|}{------ dtto ------} &
        3.2$\pm 0.05$ & 0.033$\pm 0.002$ & $160\pm 5$ & 36.38$\pm 0.15$ &
        22.5$\pm 1.0$ & 0.11$\pm 0.01$ & 1.5 / 32  \\
AG~Dra$^{e})$& 1.1 & \multicolumn{3}{c|}{------ dtto ------} &
        3.15$\pm 0.05$ & 0.032$\pm 0.002$ & $162\pm 2$ & 36.39$\pm 0.11$ &
        22.5$\pm 1.0$ & 0.11$\pm 0.01$ & 0.69 / 32  \\
\hline
\end{tabular}
\end{center}
  $^{a})$ day 292 of the X-ray observation, 
  $^{b})$ day 511 of the X-ray observation, 
  $^{c})$ on 1993/04/13 with \textsl{ROSAT}, 
  $^{d})$ as in $^{c})$, but with an atmosphere model, 
  $^{e})$ on 2004/06/15 with \textsl{XMM-Newton}, 
\normalsize
\end{table*}

\section{Discussion}

Physical parameters of SSSs, derived from the multiwavelength 
modelling the SED, are, in some cases, very different from those 
currently inferred from modelling only their X-ray fluxes. 
This is caused by the (well known) problem of mutual dependence 
between the $L_{\rm h}$, $N_{\rm H}$ and $T_{\rm h}$ parameters 
in fitting the X-ray data, because they cover only a very small 
part ($\Delta \lambda \lesssim 5$\,nm) of the total measurable 
spectrum. 
%
In the following sections I examine this problem by comparing 
both the X-ray-band and the multi-band modelling the spectra of 
SSSs in more detail. 

\subsection{Modelling only the X-ray data}

The supersoft X-ray fluxes, which are measurable by the current 
detectors, cover only the short-wavelength tail of the SSS 
spectrum, beyond its {\em unabsorbed} maximum. 
The absorption of the X-ray radiation increases markedly 
at longer wavelengths \citep[e.g. Fig~1 of][]{wilms+00}. 
These facts preclude unambiguous determination of the physical 
parameters, $L_{\rm h}$, $N_{\rm H}$ and $T_{\rm h}$, because 
they are mutually dependent, when modelling only 
the X-ray data. A larger/lower value of $L_{\rm h}$ constrains 
a larger/lower $N_{\rm H}$, but lower/larger value of $T_{\rm h}$, 
to fit satisfactorily the {\em absorbed} X-ray fluxes. 
By other words, the observed X-ray fluxes can be reproduced by 
very different sets of $L_{\rm h},~N_{\rm H},~T_{\rm h}$ 
parameters. 
An illustrative example of this effect is shown in Fig.~4 of 
\cite{ness+08}, who fitted the supersoft spectrum of the 
classical nova V723~Cas by very wide range of 
parameters, 
$\log(L_{\rm h}) = 39.33 - 36.84$ (in \es), 
$N_{\rm H} = 6.0 - 4.0 \times 10^{21}$\cmd\ and 
$T_{\rm h} = 2.72 - 3.70 \times 10^{5}$\,K. 
Other convincing examples of this effect can be found in Table~6 
of \cite{greiner+97}, Table~1 of \cite{asai+98} or in Table~2 
of \cite{sturm+11}. 

\begin{figure}
\centering
\begin{center}
\resizebox{\hsize}{!}{\includegraphics[angle=-90]{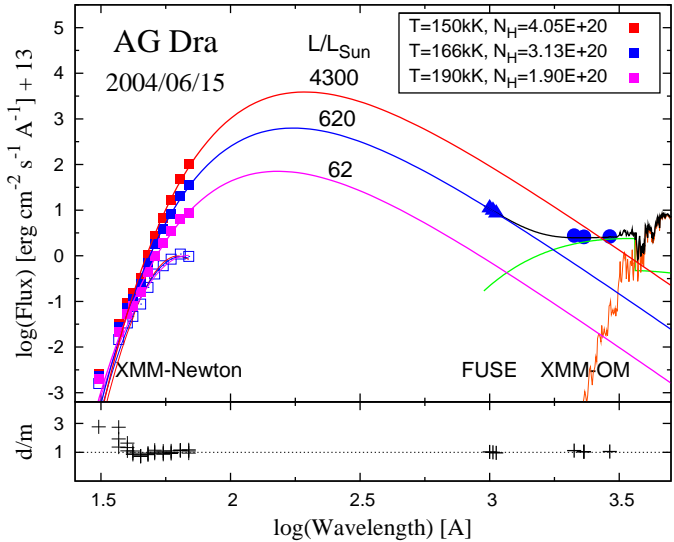}}
\end{center}
\caption[]{
Example of the mutual dependence between the $L_{\rm h}$, 
$N_{\rm H}$ and $T_{\rm h}$ parameters in modelling the 
X-ray fluxes of AG~Dra. The modelling is ambiguous, because of 
a too short wavelength range of the X-ray data (Sect.~4.1). 
To obtain the unambiguous solution, the {\em multiwavelength} 
modelling the total SED must be applied. Here, the correct 
model parameters are given by fitting the \textsl{XMM-Newton}, 
\textsl{FUSE} and \textsl{XMM-OM} fluxes (the blue line and 
points). 
          }
\label{fig:mutual}
\end{figure}

Here, I demonstrate the mutual dependence of the physical 
parameters defining the SSS radiation on the \textsl{XMM-Newton} 
spectrum of AG~Dra from 2004 June 15th (Fig.~\ref{fig:mutual}). 
The spectrum was already 
described by \cite{gon+08} and modelled by \cite{sk+09}. It covers 
the range of 31--69\,\AA. In fitting only the X-ray SED, I have 
fixed $T_{\rm h}$ and fitted the observed fluxes with the function 
(\ref{eq:fhot}) for $\theta_{\rm h}$ ($\propto L_{\rm h}$) and 
$N_{\rm H}$. 
In this way, it was possible to fit the X-ray data with 
$L_{\rm h} = 4300 - 62$\lo, 
$N_{\rm H} = 4.05 - 1.90 \times 10^{20}$\cmd\ and 
$T_{\rm h} = 1.50 - 1.90 \times 10^{5}$\,K. 
To select the correct model, we have to use fluxes from the 
far-UV, where the radiation from the SSS also dominates the 
spectrum (Figs.~\ref{fig:lin358sed},
                \ref{fig:rx0439sed}, 
                \ref{fig:v1974sed} and
                \ref{fig:agdra}). 
Here I used the \textsl{FUSE} spectrum, which was obtained on 
2004/06/24, nearly simultaneously with the \textsl{XMM-Newton} 
observations. 

It is useful to note that modelling solely the X-ray data can 
result in a misleading conclusion. For example, \cite{heise94} 
modelled the X-ray data with \textsl{ROSAT}~PSPC of the SSS 
SMP~SMC 22 (RX\,J0058.6-7136) using the blackbody and atmospheric 
model. Both models fitted well the observed X-ray fluxes (see 
their Table~1 and Fig.~2), but the bolometric luminosity of the 
blackbody ($3\times 10^{38}$\es) was more than a factor of 10 
higher than that of the model atmosphere. Therefore the authors 
concluded that fitting the WD model atmospheres yields 
a much smaller luminosity than blackbody spectra. 
However, the parameters presented in their Table~1 suggest 
that such a conclusion can be a result of the mutual 
dependence between fitting parameters, when one models only 
a very small fraction of the total spectrum. 
Therefore, I have reconstructed Fig.~2 of \cite{heise94} by 
using the parameters from their Table~1 to verify if their models 
can fit also the far-UV data. For the model atmosphere I used 
that with $T_{\rm h} = 4.5\times 10^{5}$\,K calculated by 
\cite{rauch+10} (spectrum {\small 450000-9-HHeCNONeMgSiS}, 
available at \footnote{
http://astro.uni-tuebingen.de/$^{\sim}$rauch/TMAD/TMAD.html}).
However, the ultraviolet observations with \textsl{FUSE} and 
\textsl{HST} are located far above the \cite{heise94} models 
(see Fig.~\ref{fig:heise}). 
This exposes the shortcoming of modelling the global SED of   
SSSs using only the X-ray fluxes, irrespectively of the 
model used. 
%
%

\subsection{Multiwavelength modelling the SED}

\begin{figure}
\centering
\begin{center}
\resizebox{\hsize}{!}{\includegraphics[angle=-90]{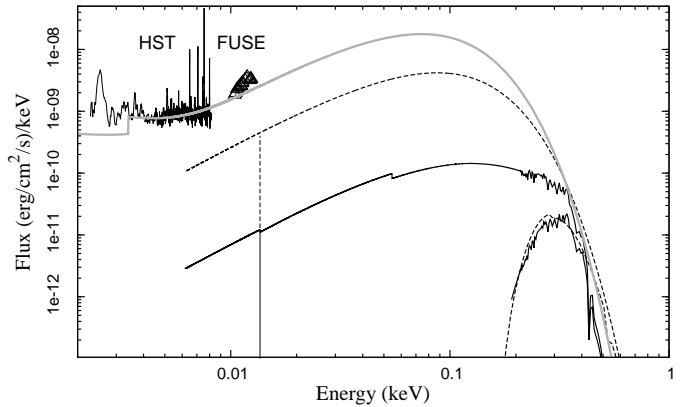}}
\end{center}
\caption[]{
The reconstructed Fig.~2 of \cite{heise94} with their blackbody
(dashed lines) and atmospheric (thick solid lines) models 
fitted to the \textsl{ROSAT} observation of SMP~SMC 22. 
Compared are measurements with \textsl{FUSE} ($\triangle$) 
and \textsl{HST} (thin line) and the multiwavelength model SED 
(gray line, Fig.~4 of Paper~II). 
          }
\label{fig:heise}
\end{figure}

To determine trustworthy physical parameters of SSSs, it is 
necessary to use fluxes from both sides of the SSS spectrum 
and to apply the multiwavelength modelling the SED. 
The far-UV fluxes put a limit for the scaling $\theta_{\rm h}$ 
in Eq.~(\ref{eq:fhot}) (i.e. the luminosity), while the X-rays 
are essential to estimate the temperature. Dereddened far-UV 
fluxes are firm (they are not dependent on the variable 
absorption by CSM), which allows us to select the correct 
model from those constrained by the X-ray data 
(see Figs.~\ref{fig:mutual} and \ref{fig:heise}). 
By other words, the multiwavelength modelling of the global SED 
of SSSs allows us to determine their $N_{\rm H}$, $L_{\rm h}$ 
and $T_{\rm h}$ parameters unambiguously, as independent 
fitting variables. 

\subsection{Comparison with previous models}
%
%

\subsubsection{LIN~358}

\begin{figure}
\centering
\begin{center}
\resizebox{\hsize}{!}{\includegraphics[angle=-90]{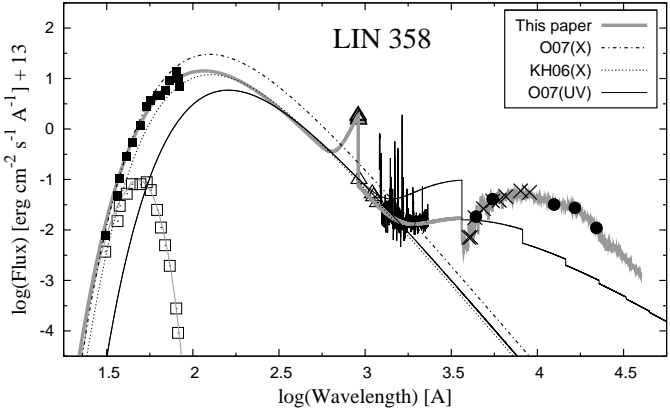}}
\end{center}
\caption[]{
Comparison of different models of the LIN~358 continuum. 
The X-ray data were fitted independently by \cite{kah+hab06} 
(KH06(X) in the legend) and \cite{orio+07} (O07(X)), who also 
modelled separately the UV observations (O07(UV) model). The 
data and multiwavelength model SED were adopted from 
Fig.~\ref{fig:lin358sed}. 
          }   
\label{fig:lin358comp}
\end{figure}

Previous blackbody models of the X-ray radiation emitted by 
LIN~358 \citep[][Sect.~3.1]{kah+hab06,orio+07} are clustered 
around the solution obtained by the multiwavelength approach 
of this paper (Fig.~\ref{fig:lin358comp}). 
\cite{orio+07} tried also to model the X-ray data with an 
atmospheric model. Their solution led to an unrealistically 
high luminosity, $L_{\rm X} \sim 8.3\times 10^{38}$\es\ within 
the 0.15--1.0\,keV range. 
They also modelled the ultraviolet \textsl{IUE} spectrum 
SWP\,49297, ignoring the X-ray and near-IR data. Their model 
($T_{\rm h} = 1.8\times 10^{5}$\,K, $R_{\rm h} = 0.127$\ro, 
$L_{\rm h} = 5.8\times 10^{37}$\es) does not reproduce 
the observations for $\lambda \gtrsim $160\,nm 
(Fig.~\ref{fig:lin358comp}). It is far above the continuum of 
the ultraviolet spectrum, because of the large emission 
measure, $EM = 1.3\times 10^{61}$\cmt\ as given by the 
quantities in their Table~2. 
In addition, such a high $EM$ cannot be generated by the hot 
stellar source in their UV model, which is not capable of 
producing the required flux of ionizing photons 
\citep[parameter $\delta > 1$, see Eq.~(21) of][]{sk05}. 

The effective temperature of the giant in LIN~358, as estimated 
by \cite{muerset+96}, is equal to that determined in this paper, 
while its radius and the luminosity are larger by a factor 
of $\sim$1.2. This is probably caused by a different approach. 
However, their parameters for the hot component, i.e. the SSS, 
are entirely different from those obtained by the 
multiwavelength modelling (Sect.~3.1, Table~2). 
The large difference in $T_{\rm h}$, $L_{\rm h}$ and 
$R_{\rm h}$ is probably a result of using the Zanstra method, 
which is very sensitive to the true level of the continuum 
adjacent to the He\I\I\ 1641\,\AA\ emission line. A faint and 
noisy UV continuum can yield a very large uncertainty in 
the Zanstra temperature. 

The multiwavelength modelling the SED of the LIN~358 spectrum 
suggests a high luminosity of its SSS, 
$(1.07 \pm 0.27)\times 10^{38}$\es. 
According to \cite{vdh+92}, the source of the radiative energy 
of 'classical' SSSs is a steady nuclear burning of the hydrogen 
rich material on the WD surface. In the case of LIN~358, a high 
mass (0.9--1.2\mo) WD accreting at $\sim 2.7\times 10^{-7}$\myr\ 
should generate the observed bolometric luminosity under 
the condition of the stable nuclear burning 
\citep[see e.g. Fig.~1 of][]{vdh11}. 
%
%

\subsubsection{RX\,J0439.8-6809}

Our solution is, within the uncertainties, identical to 
that already found by \cite{teeseling+99}, because the 
same data from both the sides of the spectrum were used. 
\cite{teeseling+99} also demonstrated that the luminosities
derived from the blackbody and atmospheric model are well  
comparable, which contradicts the previous suggestion by 
\cite{heise94} and \cite{krautter+96} (see Sect.~4.1). 

The case of RX\,J0439.8-6809 justifies the significant 
contribution of the SSS radiation to the far-UV domain 
(see Fig.~\ref{fig:rx0439sed}). 
The luminosity of RX\,J0439.8-6809 is as high as the Eddington 
limit for a 1.4\,\mo\ compact object. This makes it difficult 
to reveal the true nature of this SSS (see references in 
Sect.~3.2). 
The basic two questions, (i) what is the source of the high energy 
output, when a main-sequence donor star is not consistent with 
observations \citep[][]{teeseling+97}, and (ii) why such the 
luminous source does not generate any mass outflow, 
have not been answered yet satisfactorily. 
%
%
%
\begin{figure}
\centering
\begin{center}
\resizebox{\hsize}{!}{\includegraphics[angle=-90]{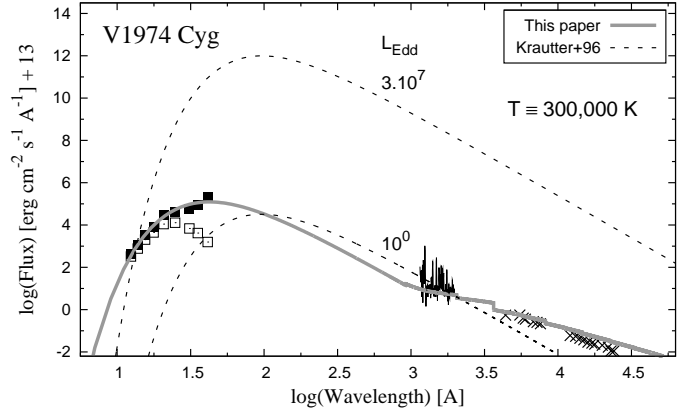}}
\end{center}
\caption[]{
Fitting the X-ray data of V1974~Cyg, \cite{krautter+96} 
obtained $T_{\rm h} = 220600 - 302000$\,K, $N_{\rm H}$ 
of several times $10^{21}$\cmd, but totally unrealistic 
$L_{\rm h}$ (upper dashed line). Therefore they adopted 
$L_{\rm h} = L_{\rm Edd}$ and $T_{\rm h} = 3\times 10^5$\,K 
(lower dashed line). 
          }   
\label{fig:v1974comp}
\end{figure}

\subsubsection{V1974~Cyg}

The multiwavelength model SED of the classical nova V1974~Cyg 
fits satisfactorily fluxes from both the Wien and the 
Rayleigh-Jeans tail of the spectrum and corresponds to 
expected physical parameters (Fig.~\ref{fig:v1974sed}, 
Table~2). The evolution during the plateau phase of the nova 
was consistent with a constant luminosity, 
$2.1-1.9(1.25\,M_{\odot}/M_{\rm WD})L_{\rm Edd}$ for the 
distance of 1.8\kpc. The multi-band model also suggested 
a constant $N_{\rm H}$, whose values, $1.8-1.7\times 10^{21}$\cmd, 
can be attributed to the interstellar quantity, because 
they are consistent with the extinction to the nova, 
$E_{\rm B-V} = 0.36$ (Sect.~3.3) according to the 
relationship, $N_{\rm H}/E_{\rm B-V} \sim 4.93\times 
10^{21}{\rm cm^{-2}mag^{-1}}$ \citep[][]{d+s94}. 
As expected, at the end of the SSS phase the multi-band model 
indicated an increase in the temperature and a decrease in 
the effective radius of the WD photosphere (Table~2). 

The large difference between the \cite{krautter+96} fitting
parameters and those derived by the {\em multiwavelength} 
approach, is caused by the mutual dependence between $N_{\rm H}$, 
$L_{\rm h}$ and $T_{\rm h}$ parameters in fitting the short 
X-ray range of fluxes. According to this effect, selection of 
a significantly lower temperature than the real one, requires 
an unrealistically high luminosity and absorbing column density 
to fit the X-ray data (see Sect.~4.1). \cite{krautter+96} 
adopted $T_{\rm h} = 3\times 10^5$\,K, which required a totally 
unrealistically high $L_{\rm h}$ and $N_{\rm H}$ to match 
the X-ray data. 
Therefore, to satisfy theoretical calculations, they scaled 
the model with the Eddington luminosity, which, however, lies 
outside observations (Fig.~\ref{fig:v1974comp}). 
Comparing the same \textsl{ROSAT} data with an atmospheric 
model calculated for a significantly higher temperature, 
\cite{balman} derived more realistic parameters (see their 
Table~1). They are comparable with those of this paper 
(Table~2). A lower effective temperature, somewhat higher 
values of $N_{\rm H}$ and a factor of $\sim 4$ lower luminosity 
in the Balman's et al. (1998) model is again a result of 
modelling solely the X-ray fluxes. 

Also in this case, modelling only the X-ray data showed that 
the problem in determining correct parameters of the SSSs 
radiation is primarily caused by using the insufficient 
wavelength range of the X-ray fluxes and not by using 
a blackbody model. 

\subsubsection{AG~Dra}

In both our examples, multiwavelength models fit well the observed 
fluxes from the supersoft X-ray to the near-IR. A higher value of 
$\chi^2_{\rm red} = 5.4$ for the 1993 model is probably caused 
by extracting the data from the figure 
\citep[Fig.~4 of][]{greiner+97}. 
The fundamental parameters of the SSS in both models are the same 
within their uncertainties. This reflects a stability of the 
energy production of the burning WD. A marginal change is 
indicated only in the hydrogen column density. The reality of 
its possible variation is supported by the variation in the 
\textsl{EM} (Table~2, Fig.~\ref{fig:agdra}), which can be caused 
by a variation in the mass loss rate from the hot star. 
The wind particles absorb the X-ray photons, but emit at 
energies for $\lambda > 912$\,\AA\ \citep[see][]{sk+09}. 
Due to a strong absorption effect to the supersoft X-rays, 
a small change in the mass loss rate can result in a measurable 
change in $N_{\rm H}$ and $EM$. 
The resulting parameters of our two models, 
 $N_{\rm H} = (2.88 \pm 0.08)\times 10^{20}$\cmd, 
 $k_{\rm n} = (4.1 \pm 0.4)\times 10^{14}$\,cm$^{-5}$ 
and 
 $N_{\rm H} = (3.15 \pm 0.05)\times 10^{20}$\cmd, 
 $k_{\rm n} = (7.6 \pm 0.7)\times 10^{14}$\,cm$^{-5}$ 
(Table~2, Eq.~(\ref{eq:fneb})) are consistent with this 
ionization/recombination process. 

\begin{figure}
\centering
\begin{center}
\resizebox{\hsize}{!}{\includegraphics[angle=-90]{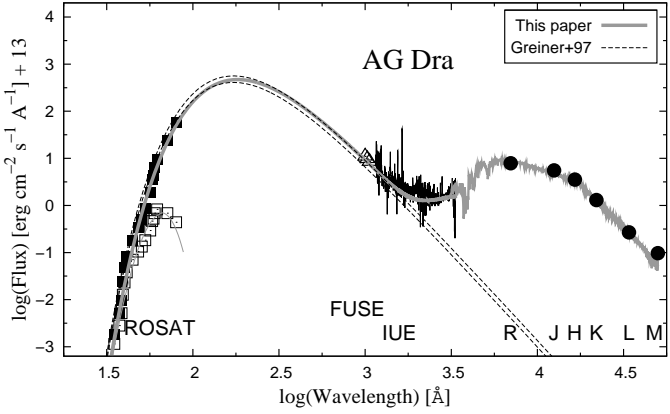}}
\end{center}
\caption[]{
Comparison of different models SED of the \textsl{ROSAT} PSPC 
observation. The gray line represents the multiwavelength 
model of this paper and the dashed lines correspond to the 
final solution of \cite{greiner+97}. 
          }   
\label{fig:AGcomp}
\end{figure}

\cite{greiner+97} fitted the \textsl{ROSAT} PSPC observations 
with three parameters ($N_{\rm H}$, Flux, $kT$) and with two 
parameters (Flux, $kT$) keeping $N_{\rm H}$ fixed to 
3.15$\times 10^{20}$\cmd. 
The former set of models always required a lower $kT$ for 
larger $N_{\rm H}$ and flux ($\propto L_{\rm h}$) than in 
the latter models (see their Table~6). Their results thus 
demonstrated the mutual dependence between the $L_{\rm h}$, 
$N_{\rm H}$ and $T_{\rm h}$ parameters 
(see Sect.~4.1, Fig.~\ref{fig:mutual}). 
However, comparing the SSS component in the AG~Dra spectrum to 
the \textsl{IUE} spectra from quiescence (see their Fig.~5), their 
resulting model for $N_{\rm H} \equiv 3.15\times 10^{20}$\cmd\ 
($kT =14.5$\,eV, $L_{\rm h} = (9.5 \pm 1.5)\times 10^{36}
(d/2.5\kpc)^2$\es) was, within uncertainties, identical in 
$L_{\rm h}$ with the multiwavelength model SED of this paper 
(Fig.~\ref{fig:AGcomp}). 
%
%
%
%
%
%
\begin{figure}
\centering
\begin{center}
\resizebox{\hsize}{!}{\includegraphics[angle=-90]{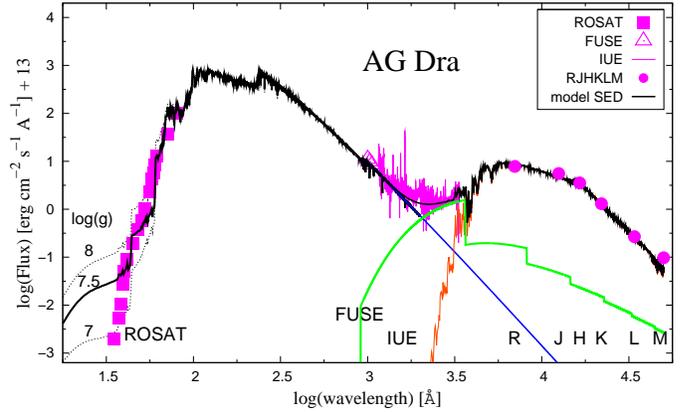}}
\end{center}
\caption[]{
Example of modelling the SED of AG~Dra using the atmosphere 
models. Physical parameters of the SSS are comparable with 
those determined from the blackbody model 
(Table~2, Sect.~4.4). 
           }
\label{fig:AGatm}
\end{figure}

\subsection{Comparison of blacbody and atmospheric model SED}

To compare a blackbody model with an atmospheric model, it is 
important that the profile of the Planck curve for a very high 
temperature is nearly identical with that given by atmospheric 
models of the hot WDs for $\lambda \gtrsim 200$\,\AA\ 
\citep[e.g. Fig.~1 in][ Fig.~\ref{fig:heise} here]{rauch03}. 
Therefore, to fit the X-ray---UV SED, one always needs to scale 
both the atmospheric and the blackbody model to the far-UV 
fluxes with a similar $\theta_{\rm h}$. This thus implies 
that also the corresponding $L_{\rm h}$ values will be 
comparable in both cases. 
In general, the atmosphere models have a steeper Wien tail 
of the spectrum than the blackbody, because of the presence 
of deep absorption lines and absorption edges seen in the 
high-resolution X-ray spectra \citep[e.g.][]{ness+11}. 
Thus to match the specific profile of the supersoft fluxes, 
a model atmosphere is required to obtain more reliable 
parameters than with a simple blackbody. 
However, the {\em multiwavelength} modelling must be used. 

Here, I demonstrate an application of modelling the SED 
with the atmospheric model for the case of AG~Dra (see 
Fig.~\ref{fig:agdra}, left panel). For the purpose of this 
illustration I used publicly available NLTE atmosphere models 
described by \cite{rauch03} and \cite{rauch+10}. I used models 
with $T_{\rm eff} = 160000$\,K, the H-Ni halo abundances and 
$\log(g) = 7-8$. Figure~\ref{fig:AGatm} shows the best 
comparison using the spectra {\small 0160000-7.00-H-Ni-halo} 
and {\small 0160000-8.00-H-Ni-halo} interpolated to 
$\log(g) \sim 7.5$. 
The model corresponds to the luminosity $L_{\rm h} \sim 630$\lo, 
$R_{\rm h}^{\rm eff} \sim 0.032$\ro\ and 
$N_{\rm H} \sim 3.2\times 10^{20}$\cmd, which are similar to 
those given by the blackbody fit (Table~2). 

In modelling the global SED by the multiwavelength approach 
with the aim to estimate the fundamental $L,R,T$ and 
$N_{\rm H}$ parameters of a SSS, selection of a model 
(blackbody or atmospheric) is not of crucial importance. 

\section{Summary}

In this paper I investigated the supersoft X-ray to near-IR SED 
of selected SSSs with the aim to determine their physical 
parameters by disentangling their composite spectra (Table~2). 
To model the global SED between 
$\sim$12\,\AA\ and $\sim$5\,$\mu$m, I used the multiwavelength 
modelling (Sect.~2). This represents the main novelty of this 
paper. The method was demonstrated on two extragalactic SSSs, 
the SyXB LIN~358 in the SMC, a bright SSS RX\,J0439.8-6809 in 
the LMC and two Galactic SSSs, the classical nova V1974~Cyg 
during its supersoft phase and the classical symbiotic star 
AG~Dra during its quiescent phase. 
The main results can be summarized as follows. 
\begin{enumerate}
\item
The models SED showed that the SSS radiation dominates also 
the far-UV domain. 
With the exception of RX\,J0439.8-6809, the modelling identified 
the {\em nebular} component of radiation, which dominates the 
spectra from the mid-UV to longer wavelengths (further 
conclusive examples are found in Paper~II of this series). 
%
\item
The multiwavelength modelling the SED overcomes the problem 
of the mutual dependence between the $L_{\rm h}$, $N_{\rm H}$ 
and $T_{\rm h}$ parameters, which arises in modelling only 
the X-ray data. The multi-band fitting procedure allows to 
determine these parameters as independent variables 
(Sects.~4.1. and 4.2., Fig.~\ref{fig:mutual}). 
\item
The physical parameters of the global SED do not depend basically 
on the model used. A blackbody or an atmospheric model yield 
a similar luminosity, because they are nearly identical in the 
profile for $\lambda \gtrsim 200$\,\AA\ and thus are equally 
scaled to the far-UV fluxes (Sects.~4.1., 4.3.2., 4.3.3. and 
4.4., Figs.~\ref{fig:heise} and \ref{fig:AGatm}). 
\item
An independent determination of the hydrogen column density 
in the ISM, based on the Rayleigh scattering the continuum 
photons around the Ly-$\alpha$ line, is suggested. Values of 
$N_{\rm H}$ obtained from modelling the SED and the Rayleigh 
attenuated continuum are consistent 
(Sects.~2.3.1. and 3.2.1., Fig.~\ref{fig:rx0439ray}). 
\item
Finally, the multiwavelength approach to modelling the SED of 
SSSs, as introduced in this paper, represents the main 
conceptional difference in comparison with previous approaches. 
\end{enumerate}

\section*{Acknowledgments}
The author thanks the anonymous referee for a detailed 
commenting on the manuscript. 
\textsl{FUSE}, \textsl{IUE} and \textsl{HST} data presented  
in this paper were obtained from the MAST. STScI is operated 
by the Association of Universities for Research in Astronomy,
Inc., under NASA contract NAS5-26555. Support for MAST for  
non-HST data is provided by the NASA Office of Space Science
via grant NNX09AF08G and by other grants and contracts.
The author thanks Rossario Gonz\'alez-Riestra for providing 
\textsl{XMM-Newton} observations of AG~Dra. 
Early versions of this work were presented at conferences 
in Brno (2009), Bowness-on-Windermere (2010), 
Vi${\tilde n}$a del Mar (2011) and Asiago (2011). 
This research has been in part supported by the project 
No. SLA/103115 of the Alexander von Humboldt foundation and 
by the Slovak Academy of Sciences under a grant VEGA 
No.~2/0002/13.

\end{document}